 \documentclass[preprint,10pt]{elsarticle} 

\usepackage[utf8]{inputenc}
\usepackage{graphicx}
\usepackage{amsmath}
\usepackage{subfig}
\usepackage{float}
\usepackage{cite}
\usepackage{amssymb}
\usepackage[symbol]{footmisc}

\usepackage[colorlinks,bookmarksopen,bookmarksnumbered,citecolor=blue,urlcolor=blue]{hyperref}
\usepackage[usenames,dvipsnames,svgnames,table]{xcolor}
\newcommand{\red}[1]{{\color{red} #1}}

\begin{document}

\begin{frontmatter}
\title{
A Discrete Droplet Method for Modelling Thin Film Flows}


\author[inst1]{Anand S Bharadwaj\footnote{Corresponding author: anand.bharadwaj@itwm.fraunhofer.de}$^,$}

\affiliation[inst1]{organization={Fraunhofer ITWM},
            addressline={Fraunhofer-Platz 1}, 
            city={Kaiserslautern},
            postcode={67663}, 
            country={Germany}}

\author[inst1]{Joerg Kuhnert}
\author[inst2,inst3]{Stéphane P.A. Bordas\footnote{stephane.bordas@alum.northwestern.edu}}
\author[inst1,inst2]{Pratik Suchde\footnote{pratik.suchde@itwm.fraunhofer.de}}

\affiliation[inst2]{organization={University of Luxembourg},
            addressline=	{2 avenue de l’universite}, 
            postcode={L-4365}, 
            state={Esch-sur-alzette},
            country={Luxembourg}}

\affiliation[inst3]{organization={Department of Medical Research, China Medical University Hospital},
            addressline=	{China Medical University},
            state={Taichung},
            country={Taiwan}}
            

\begin{abstract}

In this paper, we present a new model to simulate the formation, evolution, and break-up of a thin film of fluid flowing over a curved surface. Referred to as the discrete droplet method (DDM), the model captures the evolution of thin fluid films by tracking individual moving fluid droplets. In contrast to existing thin-film models that solve a PDE to determine the film height, here, we compute the film height by numerical integration based on the aggregation of droplets. The novelty of this approach in using droplets makes it suitable for simulating the formation of fluid films, and modelling thin film flows on partially wetted surfaces. The DDM is a Lagrangian approach, with a force balance on each droplet governing the motion, and derivatives approximated using a smoothed particle hydrodynamics (SPH) like approach. The proposed model is thoroughly validated by comparing results against analytical solutions, against the results of the shallow-water equations for thin film flow, and also against results from a full 3-D resolved Navier--Stokes model. We also present the use of the DDM on an industrial test case. The results highlight the effectiveness of the model for simulations of flows with thin films. 
\end{abstract} 
\end{frontmatter}

\section{Introduction}
Thin-film flows involve the motion of fluids in a shallow region called a film, where the height of the fluid layer is much smaller than its other dimensions. Predicting the behaviour of thin fluid films flowing over solid surfaces is essential in the development of various scientific and industrial processes, including lubrication \cite{luo1996thin}, coatings in medical applications \cite{overhoff2009use,jiang2012thin}, protective coating on lenses \cite{lee1999ftir}, solar cells \cite{tvingstedt2008trapping}, and liquid film sensors \cite{o2012review}. Another example where thin fluid film behaviour is relevant is in the automotive industry \cite{rabiei2013rainfall}. When a vehicle is moving through rain, thin water films form on the surface of the vehicle. The evolution of these rainwater films affects the performance of exterior-mounted sensors, and the aerodynamic characteristics of the vehicle, among other considerations. In this paper, we propose a novel approach to modelling thin-film flows over  general curved surfaces.

By definition, the height (also referred to as depth)  of a fluid film is a lot smaller than the other length scales of the flow, occasionally by several orders of magnitude. In the example of a vehicle moving through rain, the thickness of rainwater films is typically of the order of a few millimeters, while the width of the thin fluid layer could extend to the entire vehicle length, which is in the order of a few metres. 
When modelling these films computationally, resolving the entire fluid domain with a three dimensional discretization will lead to a restrictively large set of elements (in mesh-based methods) or points (in meshfree methods) due to the disparity in the length scales involved. This gives rise to a need for specialized thin-film models to predict fluid film behaviour at a reasonable computational cost. 

The modeling of thin film flow is a topic that has been researched for many decades (see, for example, \cite{benney1966long,ng1994roll}), and several fluid film models have been developed. One of the most prominent models used is the lubrication model or the long-wave approximation \cite{atherton1976derivation, o2015thin, myers1998thin, danov1998stability, bertozzi1996lubrication, barra2016interfacial, schwartz1995modeling}, which takes the form of a system of fourth-order PDEs. 
Another popular model is the shallow water equations \cite{fernandez2010shallow, noble2013thin, chun2017method, ren2017real}, which can be derived by depth integrating the Navier--Stokes equations. A thin-film model that has been subject to a lot of theoretical studies recently is the re-derivation of Navier--Stokes on manifolds \cite{fang2020nash,samavaki2020navier,chan2017formulation} that assumes a film of constant height.
The key commonality between all these models is that the fluid film is only discretized along the solid surface on which it is flowing, and not in the direction normal to the surface. The depth of the film, in the direction normal to the surface, is either assumed constant, or is determined by solving an additional PDE. 
In contrast, in the present work, rather than solving an additional PDE for the height function, the height of the film is built-up using a numerical integration procedure. 


The thin-film flow applications have been computed numerically using various discretization methods, ranging from finite volume methods using the volume of fluid  (VOF) approach \cite{zhao2002high,nguyen2016free,issakhov2018numerical,larmaei2010simulation}, finite element methods \cite{fries2018higher,jankuhn2018incompressible}, spectral methods \cite{gross2018hydrodynamic}, Smoothed Particle Hydrodynamics (SPH) \cite{solenthaler2011sph, chang2011numerical, wang2021thin, kordilla2013smoothed, hardi2019enhancing,hardi2020simulating}, and meshfree generalized finite difference method (GFDM) \cite{suchde2019fully,suchde2019meshfree,suchde2021meshfree}.
In the present work, we propose modelling a fluid film using a modified discrete particle model, which is commonly used for modelling small solids and aerosols \cite{lu2017assessment,zahari2018introduction,longest2011evaluation}. We represent a fluid film with a collection of discrete spherical fluid droplets. The droplets move in a Lagrangian sense, with their motion governed by a force balance that takes into account forces due to gravity, pressure (height) gradient, and viscous forces. Derivative computation, including those for the height function, are computed using a SPH-type approach \cite{liu2003smoothed}. 

Existing thin film models typically require a thin film to begin with, and they model the evolution of this film. In contrast, in the proposed discrete droplet method (DDM), the presence of an initial film is not required. We also model the formation of a film as fluid droplets impinge on a surface, making our model suitable for applications that involve dynamically forming films. Another advantage of the DDM concerns partially wetted surfaces. Existing models usually require the discretization of the entire surface over which the film flows, irrespective of the physical extent of the fluid film. This would require the allocation of unused memory in regions where the thin film does not exist, where the height is simply set to zero. In the proposed DDM, droplets also serve as numerical locations where the governing PDEs are solved. Thus, in this model, the thin film exists only in regions where the droplets are present, making the method well-suited for simulating partially wetted surfaces. 

The layout of the paper is as follows. Section \ref{sec:prelims} presents some preliminary concepts explaining the starting point of our computational framework. Section \ref{sec:model} introduces the discrete droplet method. Section \ref{sec:1d_height} presents a simplified 1-D version of the full model, which will be used in the first part of validation tests. Section \ref{sec:results} presents numerical results of the proposed model, and validations using several benchmark problems. Section \ref{sec:conclusions} presents the conclusions and directions of future work. 

\section{Preliminaries}\label{sec:prelims}

We consider a thin layer of fluid flowing over an oriented stationary surface, $\Gamma$ (in $\mathbb{R}^3$). The surface is assumed to be given by a mesh of triangles. If the surface is prescribed in other format, for example, a CAD format, then creating a surface mesh forms the first step of the numerical discretization process. Along with the surface discretization, a set of consistently oriented normals is also assumed to be known. These normals could be computed either directly from the surface mesh, or could be based on the CAD or implicit surface definition, if known. Alternately, instead of using a surface mesh as is done throughout this work, the surface may also be prescribed by a point cloud, as outlined in our earlier work \cite{suchde2021meshfree}. 

\section{The Discrete Droplet Method}\label{sec:model}

In this section, we introduce the novel discrete droplet method (DDM) for modelling thin film flows. 

\subsection{Outline}
In DDM, a fluid film is discretized as a collection of fluid droplets. The droplets are assumed to be spheres of prescribed diameters. Each droplet is represented computationally by a single point particle lying on the surface $\Gamma$. The surface normal at each droplet location is taken to be the surface normal of the element of the surface mesh on which the droplet lies. Similar to other thin film models, this is a pseudo $2$-dimensional model where the motion of the droplets occurs along the surface $\Gamma$, and the computed height of the film at each droplet location accounts for the third dimension.

Derivatives are computing using a SPH-like approach. For this, for each droplet, a neighbourhood or support must be determined. This neighbourhood is computed based on proximity to other droplets. For a droplet $i$, the neighbourhood $S_i$ is given by all droplets within a distance $h$ of $i$. All distance computations required are done as Euclidean distances in $\mathbb{R}^3$, and not along $\Gamma$. To compute the neighbourhoods efficiently, we follow neighbour search algorithms commonly used in meshfree methods \cite{dominguez2011neighbour,drumm2008finite}. The fluid droplets move in a Lagrangian sense with the fluid velocity. Since droplets are moving, the neighbourhood of the droplets could change between time-steps. To save computational time, rather than recomputing the neighbourhood in every time step, we recompute every $3-5$ time steps, depending on the application.

 The velocity equation that governs the motion of the droplets is first discussed. This is followed by a discussion on droplet movement and the computation of film height and its derivatives, which appear in the velocity equation. Subsequently, solution stabilization, time integration summary, initial conditions for height, and a simplified free-flight droplet model are discussed.   

\subsection{Governing Equation}
The motion of a droplet on a surface is governed by forces of gravity, pressure and viscosity. The velocity of a droplet on a general surface is given by a force balance on each droplet, and can also considered as a variation of the Cauchy momentum equation.
\begin{equation}
\label{eq:2d_momentum}
\frac{d\vec{V}_{\text{drop}}}{dt} = \frac{\eta_{\text{drop}}(\vec{V}_{\text{s}}-\vec{V}_{\text{drop}})}{\rho_{\text{drop}}H_{\text{film}}^2} + \hat{g} - \frac{\nabla P}{\rho_{\text{drop}}}
\end{equation}
where \\
the subscripts \lq s' denotes the surface,\\ 
$\vec{V}_{\text{drop}}$ is the velocity of the droplet on the surface such that $\vec{V}_{\text{drop}} \cdot \vec{n} = 0$, $\vec{n}$ being the unit normal to the surface,\\
$\vec{V}_{\text{s}}$ is the velocity of the surface,\\
$\eta_{\text{drop}}$ is the viscosity of the droplet,\\
$\hat{g} = \vec{g} - (\vec{g} \cdot \vec{n})\vec{n}$ is the tangential gravitational acceleration, with $\vec{g}$ is the acceleration due to gravity. \\
Furthermore, $H_{\text{film}}$ is the reconstructed film height,\\
$P = P_{hyd} + P_{\sigma}$ is the sum of the hydrostatic pressure due to the height of the film, $P_{hyd}=\rho_{\text{drop}} (\vec{g}.\hat{n}) H_{\text{film}}$ and pressure due to surface tension, $P_{\sigma} = \sigma(\kappa_f + \kappa_s)$. Here, $\kappa_f$ and $\kappa_s$ are the curvatures of the film and the surface respectively. 

\begin{figure}
\centering
\includegraphics[trim={0 2cm 0 10cm},width=0.9\textwidth]{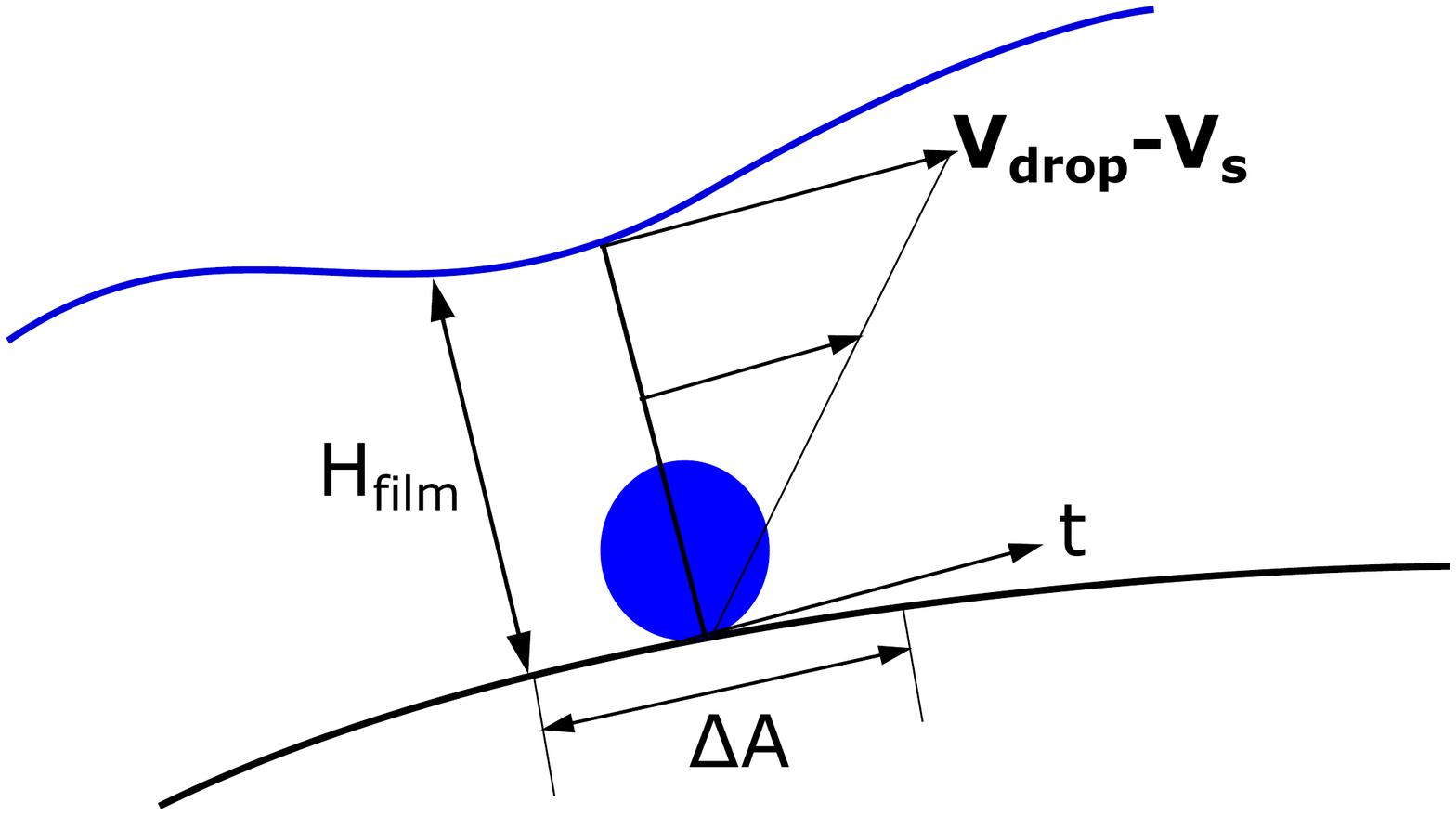}
\caption{Illustration of a droplet used to represent a portion of a thin fluid film: the reconstructed film height ($H_{\text{film}}$), velocity profile and the tangent vector $\vec{t}$ are shown.}
\label{f:viscforce}
\end{figure}

While the pressure and gravity terms appear in their standard form in Eq.~\eqref{eq:2d_momentum}, the viscous term is formulated specifically for this model. As per Newton's law, the effect of the viscous force on the droplet, over an area $\Delta A$ of the surface, may be written as
\begin{equation}
\label{eq:visc_force}
\vec{F}_{v} =  \Delta m \frac{d\vec{V}_\text{drop}}{dt} = -{\eta_{\text{drop}}} \frac{\partial \|\vec{V}_\text{drop}\|}{\partial n} \bigg \rvert_{n=0} \vec{t} \Delta A
\end{equation}
In the above equation, $\Delta m$, $n$, $\|\vec{V}_\text{drop}\|$  and $\vec{t}$ are the mass of the droplet,  surface normal coordinate, $L_2$ norm of the droplet velocity and the unit tangent vector respectively. The preceding negative sign on the RHS implies that it acts as a decelerating force. As seen in Figure \ref{f:viscforce}, at the location of a droplet, a linear profile for the velocity is assumed such that 
\begin{equation}
\frac{\partial \|\vec{V}_\text{drop}\|}{\partial n} \bigg \rvert_{n=0} \approx \frac{\|\vec{V}_{\text{drop}}-\vec{V}_s\|}{H_{\text{film}}}
\end{equation}
Here, we assume that at the surface, the fluid layer has no relative velocity, and moves with the surface velocity. 
Therefore, Eq.~\eqref{eq:visc_force} becomes
\begin{equation}
\vec{F}_{v} = \rho_{\text{drop}} \Delta V \frac{d\vec{V}_{\text{drop}}}{dt} = -\eta_{\text{drop}} \frac{\|\vec{V}_{\text{drop}}-\vec{V}_s\|}{H_{\text{film}}}\vec{t} \Delta A 
\end{equation} 
where $\Delta V$ is the volume of the droplet. On rearranging the terms and writing the unit tangent vector $\vec{t}$ as $(\vec{V}_{\text{drop}}-\vec{V}_s)/\|\vec{V}_{\text{drop}}-\vec{V}_s\|$, we get
\begin{equation}
\frac{d\vec{V}_{\text{drop}}}{dt} = -\frac{\eta_{\text{drop}}}{\rho_{\text{drop}}} \frac{\|\vec{V}_{\text{drop}}-\vec{V}_s\|}{H_{\text{film}}} \frac{\vec{V}_{\text{drop}}-\vec{V}_s}{\|\vec{V}_{\text{drop}}-\vec{V}_s\|} \frac{\Delta A}{ \Delta V} 
\end{equation} 
Setting $\Delta A/\Delta V$ as $1/H_{\text{film}}$, 
\begin{equation}
\frac{d\vec{V}_{\text{drop}}}{dt} = -\frac{\eta_{\text{drop}}}{\rho_{\text{drop}}} \frac{\vec{V}_{\text{drop}}-\vec{V}_s}{H_{\text{film}}^2}  =  \frac{\eta_{\text{drop}}}{\rho_{\text{drop}}} \frac{\vec{V}_{s}-\vec{V}_{\text{drop}}}{H_{\text{film}}^2}
\end{equation} 
which is the viscous force as seen in the first term on the RHS of Eq.~\eqref{eq:2d_momentum}. 

Equation ~\eqref{eq:2d_momentum} is used to solve for the velocity and thereby, also determines the position of each droplet.
The height function and its derivatives are computed by the SPH-formulation and are described below. Henceforth, $H_{\text{film}}$ is referred to as $H$ for brevity. We note here the difference between $H$ which denotes the film height, and $h$, which refers to smoothing length or interaction radius that determines the droplet neighbourhoods. 

\subsection{Droplet movement}
\label{sec:move}
Droplets always move in a direction tangential to the surface i.e. the surface normal component of the velocity is always zero. The position of the droplet is updated based on a second-order method \cite{suchde2018point}. Time integration from time level $t^{(n)}$ to $t^{(n+1)}$ begins with updating the droplet positions as
\begin{equation}
\label{Eq:Move}
\Delta \vec{x} = \vec{V}_{\text{drop}}^{(n)} \Delta t + \frac{1}{2} \frac{\vec{V}_{\text{drop}}^{(n)}-\vec{V}_{\text{drop}}^{(n-1)}}{\Delta t} \Delta t^2
\end{equation}  
Note that the new velocity $\vec{V}_{\text{drop}}^{(n+1)}$ will be computed on the update droplet positions is not yet known when the droplets are advected. To ensure that a droplet remains on the surface during the movement process, it is advected along the plane of the surface element it lies on. When the position of the droplet crosses the bounds of this element, it is projected back on to the surface defined by the adjacent surface element.

\subsection{Film height computation} 
\label{sec:h_computation}

\begin{figure}
    \centering
    \includegraphics[width=0.75\textwidth]{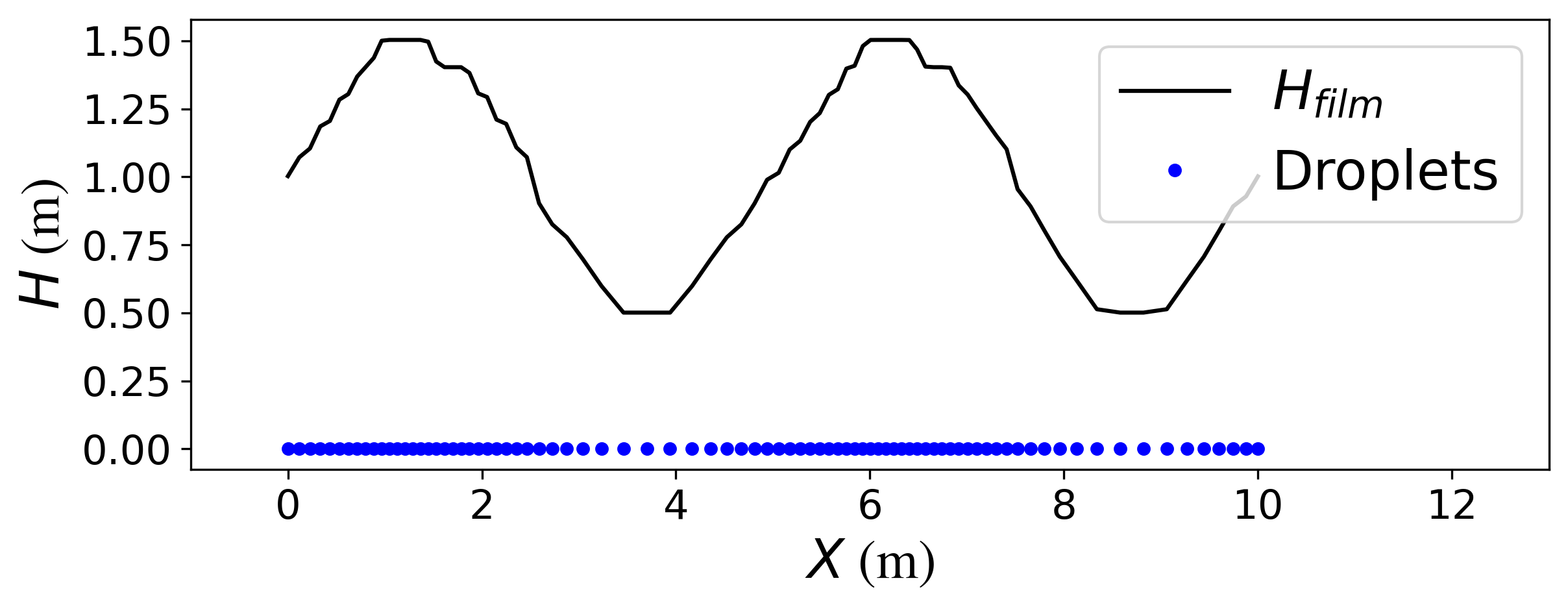}
    \caption{Effect of droplet distribution on the reconstructed film height $H_{\text{film}}$. The blue points show the droplets lying on a flat 1-D surface, while the black line shows the computed height function. Regions with a higher droplets density represent a higher film height.}
    \label{fig:droplet_density}
\end{figure}

While most thin film flow models solve an additional PDE for the height function, we instead build up the height function using a numerical integration procedure based on the collection of droplets. A larger number of droplets within the same area on the surface would mean a larger film height. This is illustrated in Figure \ref{fig:droplet_density}. 

We adopt an SPH-like formulations for numerically integrating the height function and estimating its derivatives. 
A general scalar-valued function $\phi(\vec{x})$ in 2-D may be approximated using an SPH formulation as
\begin{align}
    \phi(\vec{x}_i) &= \int \int_{\mathbb{R}^2} W(\vec{x}-\vec{x}_i) \phi(\vec{x}) dA \\
    &\approx \sum_{j\in S_i} W(\vec{x}_j-\vec{x}_i) \phi(\vec{x}_j)\Delta A_j
    \label{eq:height_2d}
\end{align}
$\Delta A$ is an elemental area in $\mathbb{R}^2$ and the summation is computed over the neighbourhood $S_i$ of droplet $i$. The compactly supported kernel function $W(\vec{x}_j-\vec{x}_i)$ is chosen as 
\begin{align}
   W(\vec{x}_j-\vec{x}_i) &= \frac{\alpha}{\pi h^2} \exp\left[ -\alpha \frac{\|\vec{x}_j-\vec{x}_i\|^2}{h^2}\right] \\
   &= \frac{C^*}{h^2} \exp\left[ -\alpha \frac{\|\vec{x}_j-\vec{x}_i\|^2}{h^2}\right]  \label{eq:2d_weight}
\end{align}
where the coefficient of the exponential term is determined by imposing the condition
\begin{equation}
    \int \int_{\mathbb{R}^2}  W(\vec{x}-\vec{x}_i) dA = 1
    \label{eq:consistency_condition}
\end{equation}
The derivation of this term is explained in \ref{sec:app1}.
Substituting Eq.~\eqref{eq:2d_weight} in Eq.~\eqref{eq:height_2d}, we get
\begin{equation}
    \phi(\vec{x}_i) = \sum_{j \in S_i}  \frac{C^*}{h_j^2} \exp\left[ -\alpha \frac{\|\vec{x}_j-\vec{x}_i\|^2}{h_j^2}\right]  \phi(\vec{x}_j)\Delta A_j
    \label{eq:SPH_weight}
\end{equation}
Since the function of interest here is the film height, we set $\phi = H$.
\begin{align}
       H(\vec{x}_i) &= \sum_{j \in S_i}  \frac{C^*}{h_j^2} \exp\left[ -\alpha \frac{\|\vec{x}_j-\vec{x}_i\|^2}{h_j^2}\right]  H(\vec{x}_j)\Delta A_j \\
       &= \sum_{j \in S_i}  \frac{C^*}{h_j^2} \exp\left[ -\alpha \frac{\|\vec{x}_j-\vec{x}_i\|^2}{h_j^2}\right] V_j
       \label{eq:height_2dd}
\end{align}
where $V_j$ is the volume of droplet $j$. We note that the expression for the height in Eq.~\eqref{eq:height_2dd} can also be derived using the argument of volume conservation, as explained in \ref{sec:app2}. Since droplets are assumed to be spherical, the volume can be computed as $V_j= \frac{1}{6}\pi d_j^3$, for diameter $d_j$.

This formulation for numerically integrating the height function can be interpreted as a smearing out of the volume of each droplet in a Gaussian shape across its neighbourhood. The height of the film at any location is computed by summing the Gaussian functions of droplets in the neighbourhood of that location. The mass of the fluid remains conserved since the number of droplets remains the same throughout the simulation, unless droplet sources/sinks are specifically defined. As the droplets advect to new locations the height function at each droplet location evolves with changing neighbourhoods. 

\subsection{Derivative computation}
\label{sec:derivatives}

In Eq.~\eqref{eq:2d_momentum}, derivative computation is only needed for the pressure turn. The gradient of the hydrostatic pressure requires the gradient of the film height, while the gradient of the surface tension pressure requires curvature computation, and thus, the second derivative of the film height.

Function derivatives are also computed using an SPH-like formulation. For a scalar-valued function $\phi$ as given in Eq.~\eqref{eq:height_2d}, it's derivative can be approximated as 
\begin{equation}
    \nabla \phi (\vec{x}_i) = -\sum_{j \in S_i} \nabla W_{ij} \phi(\vec{x}_j) \Delta A_j
\end{equation}
where $W_{ij}=W(\vec{x}_j-\vec{x}_i)$. Substituting the gradient of the kernel function, obtained from chain rule, 
\begin{equation}
       \nabla \phi (\vec{x}_i) = -\sum_{j \in S_i} \left(-\frac{2\alpha}{h_j^2}\right)(\vec{x}_j-\vec{x}_i)W_{ij} \phi(\vec{x}_j) \Delta A_j 
\end{equation}
Therefore, for the derivative of the height function,
\begin{eqnarray}
       \nabla H (\vec{x}_i) &=& -\sum_{j \in S_i} \left(-\frac{2\alpha}{h_j^2}\right)(\vec{x}_j-\vec{x}_i)W_{ij} H(\vec{x}_j) \Delta A_j  \\
                                       &=& -\sum_{j \in S_i} \left(-\frac{2\alpha}{h_j^2}\right)(\vec{x}_j-\vec{x}_i)W_{ij} V_j
\end{eqnarray}

\subsection{Stabilization using velocity smoothing}
\label{sec:smooth}

The solution from the DDM is smoothed at every time step owing to numerical instabilities due to the hyperbolic nature of the governing PDE. First, Eq.~\eqref{eq:2d_momentum} is solved to get an intermediate velocity, $\vec{V}_{\text{drop}}^*$. Subsequently, a smoothed velocity is computed as
\begin{equation}
    \vec{V}^s_{\text{drop,}i} = \frac{\sum _j w_{ij} \vec{V}^*_{\text{drop,}j}}{\sum _j w_{ij}}
\end{equation}
The smoothing function $w$ in the above equation is chosen as a Gaussian 
\begin{equation}
w_{ij} = \frac{1}{l\sqrt{\pi}} \exp\left[-\left(\frac{|\vec{x}_j-\vec{x}_i|}{l}\right)^2\right]
\end{equation}
where $\vec{x}_i$ and $\vec{x}_j$ are the position vectors of droplets $i$ and $j$ respectively. The smoothing length, $l$, is at maximum the interaction radius $h$. 
The final velocity is computed as 
\begin{equation}
\label{eq:smoothing}
    \vec{V}_{\text{drop,}i}^{(n+1)} = \omega \vec{V}^s_{\text{drop,}i} + (1-\omega)\vec{V}^*_{\text{drop,}i}
\end{equation}
In the above equation, the value of $\omega$ ($\in [0,1]$) determines the extent of smoothing used in the solution. Thus, the final velocity is not a true smoothed version of the velocity computed from the time integration, rather it is an interpolation between the computed velocity and its smoothed counterpart. While the last test case of this paper does not use smoothing, all the other test cases use $\omega=0.1$, which we empirically observe is sufficient to produce stable results.

An alternative approach for stabilization is to add an artificial viscous term. In this approach, an artificial viscous force can be added to the velocity equation Eq.~\eqref{eq:2d_momentum}
\begin{equation}
F_v^{\text{art}} = \frac{\eta_{\text{art}}}{\rho_{\text{drop}}} \nabla^2 _t\vec{V}_{\text{drop}}
\end{equation} 
In the above equation, $\nabla^2_t$ is the Laplacian operator in the tangent plane of the surface and $\eta_{\text{art}}$ is the artificial viscosity. Our numerical simulations suggest that both methods result in a similar stabilizing effect. However, when using the artificial viscosity approach, the choice of the artificial viscosity was highly application dependent. In contrast, a largely application independent value of $\omega=0.1$ produced stable results for all the test cases considered in the first approach. Thus, the artificial viscosity method is \emph{not} used, and all numerical results presented below use the stabilization approach of Eq.~\eqref{eq:smoothing}. 

\subsection{Time Integration Summary}
The time integration procedure to model the evolution of droplets in the thin fluid layer can be summarised as follows
\begin{enumerate}
    \item The first step is the Lagrangian movement, according to Eq.~\eqref{Eq:Move}.
    \item Computation of the height function at the new droplet locations, according to Eq.~\eqref{eq:height_2d}. 
    \item Computation of height derivatives, as laid out in Section~\ref{sec:derivatives}.
    \item Explicit time integration of the velocity equation, Eq.~\ref{eq:2d_momentum}, to produce an intermediate velocity. 
    \item Smoothing of velocity to produce the final velocity, as explained in Section~\ref{sec:smooth}. 
\end{enumerate}

\subsection{Initial condition for the film height}
In the DDM, the droplet distribution on the surface determines the height function, as explained above. Therefore, to achieve a specific initial condition for the height function, a droplet distribution is needed at the beginning of the simulation, that would match the given initial condition. In other words, the height computed using Eq.~\eqref{eq:height_2dd} with the initial droplet distribution, must satisfy the required initial condition. Thus, we must solve an inverse problem to find the droplet distribution for a given initial height function. The solution procedure for this inverse problem is described in detail in the next section for a 1 dimensional simplification of the DDM. 

In contrast, in existing thin film models which solve an addition PDE for the height function, no extra work is needed to enforce the required initial condition. This forms a drawback of the proposed method. However, we emphasize that the primary motivation behind the development of the present model is applications where the formation of a fluid film is as important to model as the evolution of the film. In these application, there is no initial fluid film, and thus, there is no need to solve the inverse problem. 

\subsection{Free-flight model}
\label{sec:freeflight}
To accurately capture the formation of a fluid film, the DDM must also include a model for capturing fluid flow before it hits the solid surface over which the film will form. Since the focus of this work is modelling thin film flow, we only consider a very simple first approximation for modelling the linked bulk (3-D) flow. The coupling of the introduced thin film flow model with full 3-D flow models forms an extension of this work, and will not be covered here. 

Consider a fluid droplet of diameter $d$ in free flight before it strikes a solid surface. The droplet is assumed to remain spherical, and its motion is governed by a simplified force balance
\begin{equation}
    \frac{d \vec{V}_{\text{drop}}}{dt} = \vec{g}_{\text{eff}}
\end{equation}
Here, $ \vec{V}_{\text{drop}}$ is the velocity of the droplet in free flight and $\vec{g}_{\text{eff}}$ is the effective acceleration due to gravity, and any other external forces. Once a free-flight droplets hits a solid surface, we treat it as part of a thin film, as described earlier in this section. 

\subsection{Droplets leaving the surface}
\label{sec:Leave}
The criteria for a droplet to stay on the surface is that the net normal force acting on it should be pointing in the direction of the inward surface normal. In such a case, the droplet remains a part of the thin film. When the net normal force acts in the direction of the outward surface normal, the droplet leaves the thin film and becomes a free-flight droplet, as described in Sec. \ref{sec:freeflight}. 

\section{Simplified discrete droplet method in 1-D} \label{sec:1d_height}

Before considering the validation process of the thin-film DDM developed in this work, we first perform a validation of a simplified version of the DDM which restricts flow to a one-dimensional flat surface. The verification and validation of both the simplified model and the full model are presented in Section~\ref{sec:results}. We now introduce the simplifications of the model introduced in Section~\ref{sec:model} to 1-D domains. 

A simplified version of the velocity equation in the Lagrangian form for a flat 1-D domain can be derived as a special case of Eq.~\eqref{eq:2d_momentum}, by setting all terms to zero except the hydrostatic pressure term, which leads to a height derivative 
\begin{equation}
    \frac{DV_{\text{drop}}}{Dt} + {g} \frac{\partial H}{\partial x} = 0
    \label{eq:SWE_mom}
\end{equation}
This equation can also be derived from the momentum conservation equation of the 1-D inviscid shallow-water equation for a flat domain.  

At a droplet $i$, the height derivative is used to determine the velocity, and subsequently, the position of the droplet is calculated at the next time step, as shown in Eqs.~\eqref{eq:vel} and~\eqref{eq:pos}.
\begin{equation}
    V_{\text{drop,}i}^{(n+1)} = V_{\text{drop,}i}^{(n)} - \Delta t g \frac{\partial H}{\partial x} \bigg \rvert_i^{(n)}
    \label{eq:vel}
\end{equation}
\begin{equation}
    x_i^{(n+2)} = x_i^{(n+1)} + V_{\text{drop,}i}^{(n+1)} \Delta t
    \label{eq:pos}
\end{equation}

Each droplet is associated with a kernel function given by a Gaussian.
\begin{equation}
    W(x) =\frac{\sqrt {\pi} d^2}{4h} \exp\left[-\left(\frac{x}{h}\right)^2\right] 
\end{equation}
where $d$ is the droplet diameter and $h$ is the smoothing length. This expression for height function in 1-D is determined in a similar manner to the pseudo 2-D full model, by imposing the condition as in Eq. \ref{eq:consistency_condition}. For more details, see \ref{sec:app1}.

Using the above result, the height computation at a  droplet location $x_i$, in 1-D, is given by the summation 
\begin{equation}
    H_i = \sum_{j \in S_i} \frac{\sqrt {\pi} d_j^2}{4h} \exp\left[-\left(\frac{x_j-x_i}{h_j}\right)^2\right] 
    \label{eq:height}
\end{equation}
where \\
$H_i$ is the height of the film at a droplet $i$, \\
$j$ is the summation index for the droplets in the neighbourhood of droplet $i$, \\
$d_j$ is the  diameter of droplet $j$, \\
$h$ is the smoothing length,\\
and $x_i-x_j$ is the distance between droplets $i$ and $j$.

The height derivative required in Eq.~\eqref{eq:SWE_mom} is computed using the chain rule as 
\begin{equation}
    \frac{\partial H}{\partial x} \bigg \rvert_i = -\sum_{j=1}^N \frac{\sqrt {\pi} d_j^2}{4h_j} \exp\left[-\left(\frac{x_j-x_i}{h_j}\right)^2\right]   \left( \frac{-2(x_j-x_i)}{h_j^2}\right)
    \label{Eq:height_derivative}
\end{equation}


\subsection{Initial droplet distribution in 1-D}\label{method:initial}
The method adopted to determine a droplet distribution that matches a specific initial condition for the film height is described here. Although this idea may be easily extended to 2-D, it is described here in 1-D because it has been used in this work only for the 1-D test cases. 

\begin{figure}
    \centering
    \subfloat[Step 1: Uniform initial distribution of droplets, with their initial supports marked.]{\includegraphics[width=0.45\textwidth]{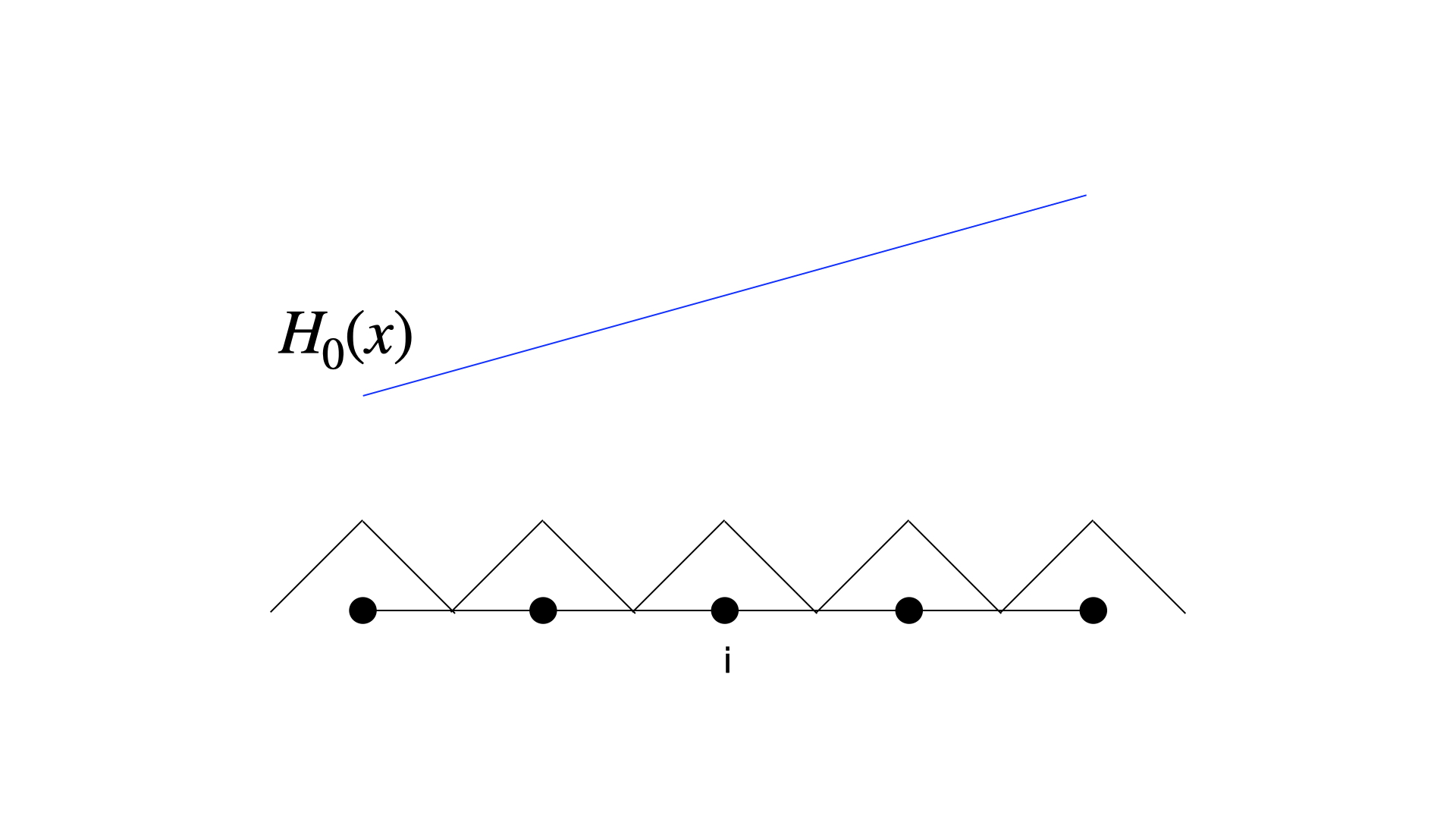}}
    \phantom{abc}
    \subfloat[Step 2: Addition of droplets within every support to match the given height at that location.]{\includegraphics[width=0.45\textwidth,trim={0 10cm 0 0}]{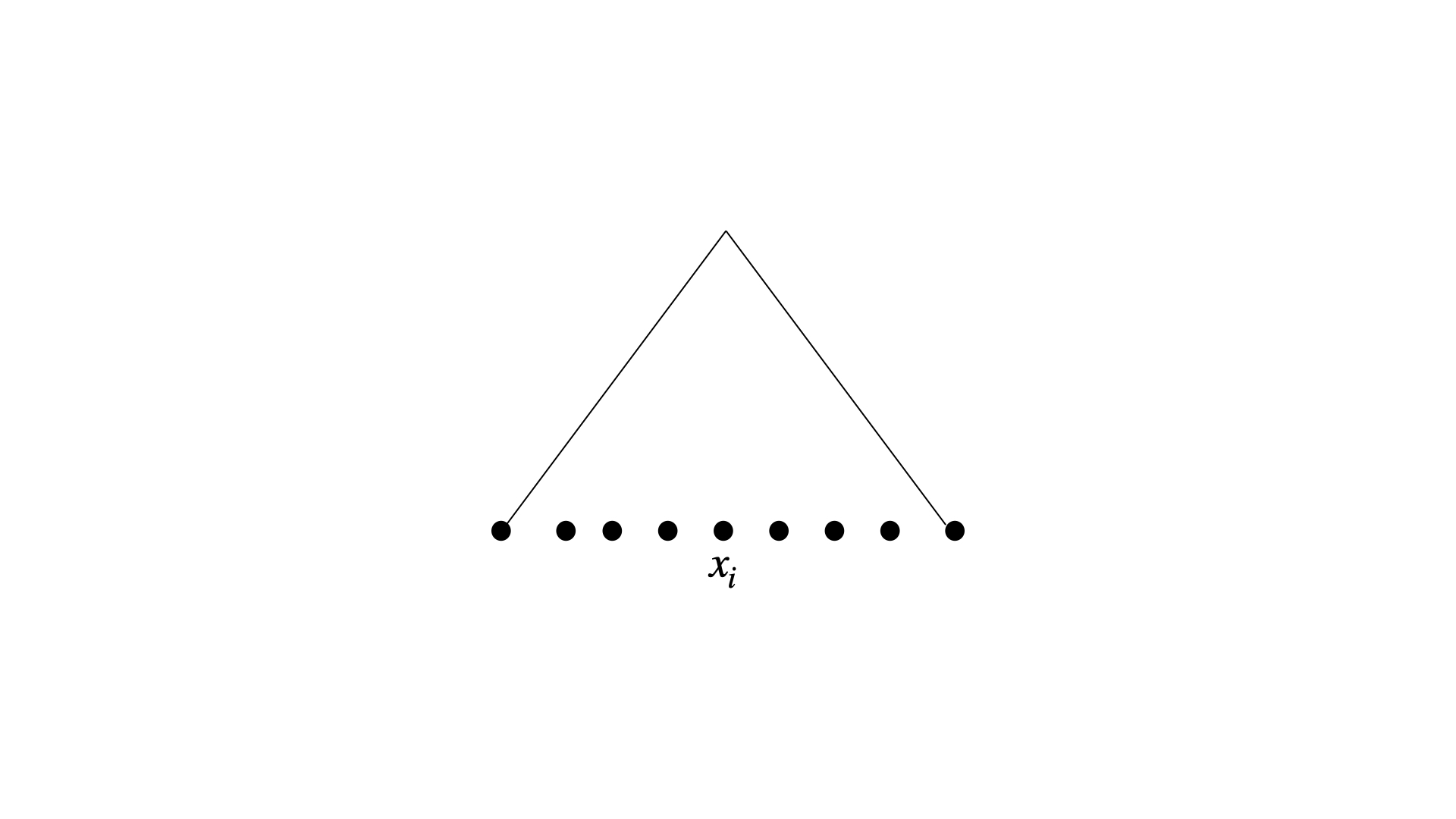}}
    \caption{Determining the initial droplet distribution condition in the thin-film DDM for a given initial condition $H_0$ of the height function: A 1-D illustration.}
    \label{fig:init}
\end{figure}

As seen in Eq.~\eqref{eq:height}, the height function is dependent on the distribution of droplets in the computational domain. In order to match an initial condition for the height function, the corresponding droplet distribution needs to be obtained at the start of the computation. Towards this end, for each droplet $i$, a triangular hat function with a predetermined support centred at the droplet is considered. As a first step, the domain is sparsely initialized with droplets such that no two droplets have an overlap in the support of their respective hat functions, as shown in Figure \ref{fig:init}(a). 

Subsequently, additional droplets are added within the support of each droplet $i$ such that the height at the location of droplet $i$ matches the height as needed by the initial condition $H_0(x)$. The additional droplets are added uniformly within the support of the $i^{th}$ droplet, with the number of droplets $N_{new}$ determined by the relation
\begin{equation}
    H_0(x_i) = \frac{(N_{new}-1)}{2}H_g
\end{equation}
Here, $H_g$ is the maximum height of the Gaussian function associated with the droplet. Figure \ref{fig:init}(b) shows the additional droplets added within the support to meet the required height initial condition. 

This procedure works well for smooth initial conditions. For discontinuous initial conditions, a small region with large gradients is inevitable to match the jumps due to the discontinuities. Additionally, at the boundaries of the domain, this procedure is bound to give rise to errors, unless ghost/mirror points are used. 


\section{Results} \label{sec:results}
In this section, a thorough verification and validation of the DDM is presented. The results of the DDM are compared with the solutions obtained from the shallow-water model, 3-D Navier--Stokes equations and analytical solutions. First, comparisons of the simplified 1-D model are presented, followed by comparisons of the full model. A list of test cases is tabulated in Table \ref{t:test_cases}, and are presented in order of increasing complexity. The test cases presented in this paper do not consider forces due to surface tension.

In existing thin-film models, an initial thin film is required and consequently, for purposes of verification and validation it is necessary to consider such test cases. Thus, in test cases 1-5, an initial thin film is used in the simulation, which is not the case in the final test case. The final test case is presented to emphasize the category of problems this model is targeted to solve. 

The presented method and algorithms of the thin-film DDM are implemented in the in-house software suite MESHFREE\footnote{meshfree.eu}. All simulations are run serially, except the last test case where an MPI-based parallelization is used.

\begin{table}
\centering
\caption{List of verification and validation test cases used.}
\label{t:test_cases}
\scalebox{0.7}{
\begin{tabular}{cccc}
\hline \hline
Case No. & Test case & Dimension of surface & Remarks \\ \hline
1& Water Hump & 1-D & Comparison against shallow-water \\  
2& Dam-break & 1-D & Comparison against shallow-water \\
3& Sliding thin film & 2-D (flat) & Comparison against analytical solution \\
4& Hemispherical bowl & 2-D (curved) & Convergence w.r.t droplet diameter \\
5& Asymmetric dam-break & 2-D (flat) & Comparison against 3-D Navier--Stokes \\
6& Flow over a spoiler & 2-D (curved) & Comparison against 3-D Navier--Stokes \\ \hline \hline
\end{tabular}}
\end{table}
\subsection{Water hump}
We start by considering flow on a 1-D flat domain $0 \le x \le 10$. In this test case \cite{leveque2002finite}, a hump of water in the domain collapses and two waves travel from the center of the domain towards the boundaries. The acceleration due to gravity $g$ is taken to be unity. The initial condition for the height and velocity is given as
\begin{eqnarray}
    H(x) &=& 1 + \frac{2}{5} \exp[-5(x-5)^2]\label{init_h} \\
    V_{\text{drop}}(x) &=& 0 \label{init_u}
\end{eqnarray}

The initial droplet distribution to satisfy the initial condition is obtained using the method explained in Sec \ref{method:initial}. In this test case, the comparison of the solutions of DDM against that of a finite-volume solution to the shallow-water equations is presented.

\begin{figure}
    \centering
    \subfloat[$t=0.5$s]{\includegraphics[width=2.5in]{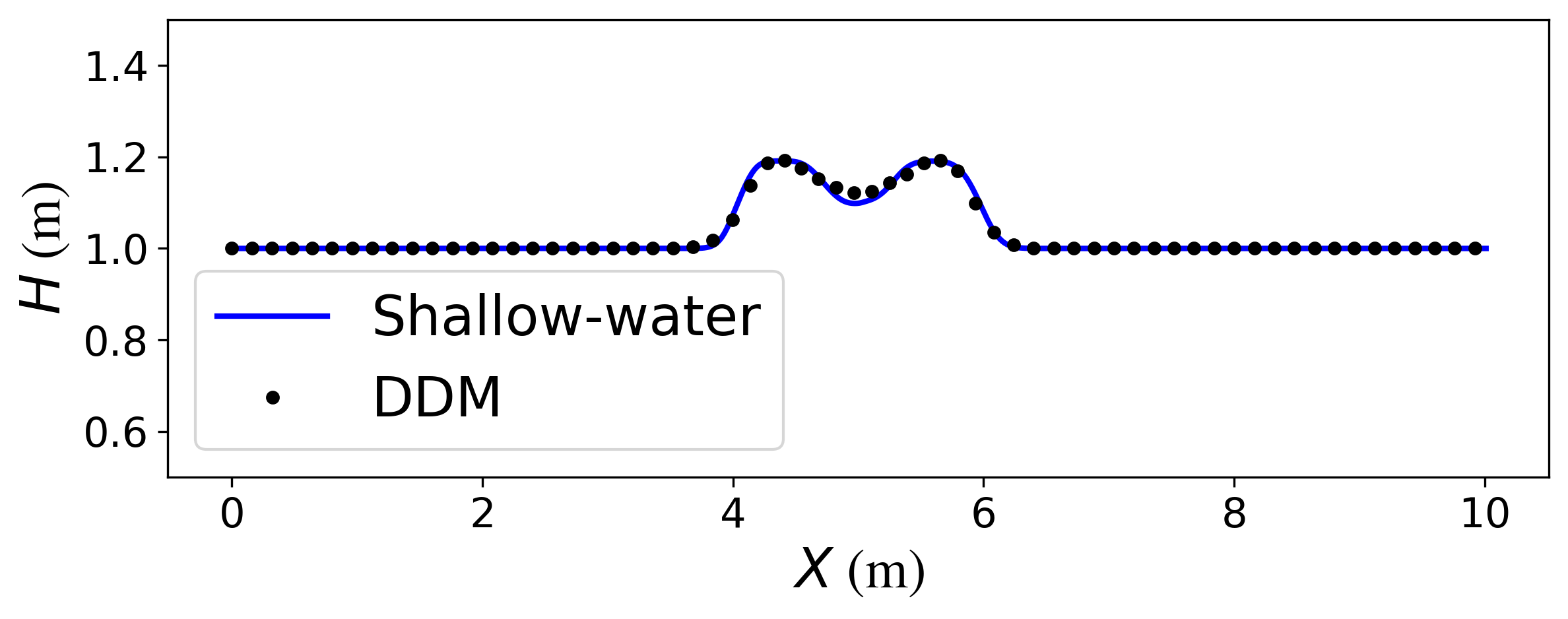}} 
    \subfloat[$t=1.5$s]{\includegraphics[width=2.5in]{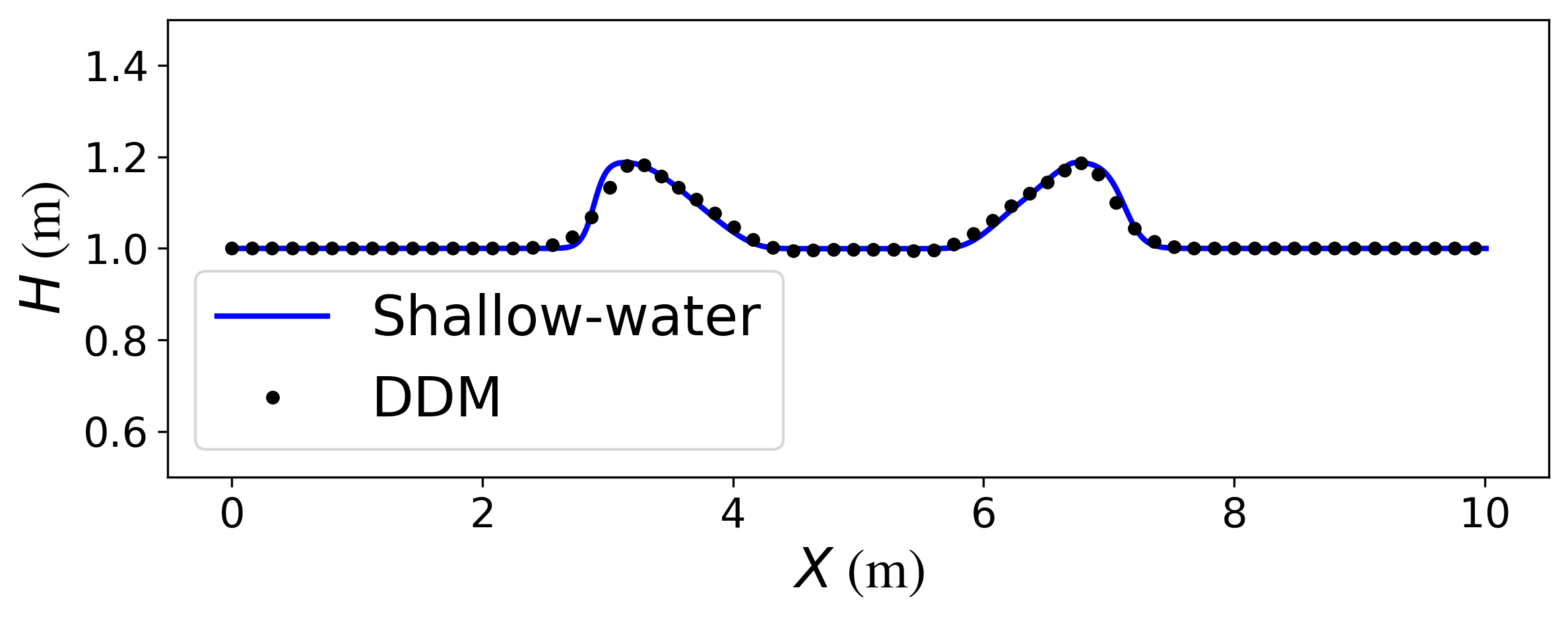}} \\
    \subfloat[$t=2.5$s]{\includegraphics[width=2.5in]{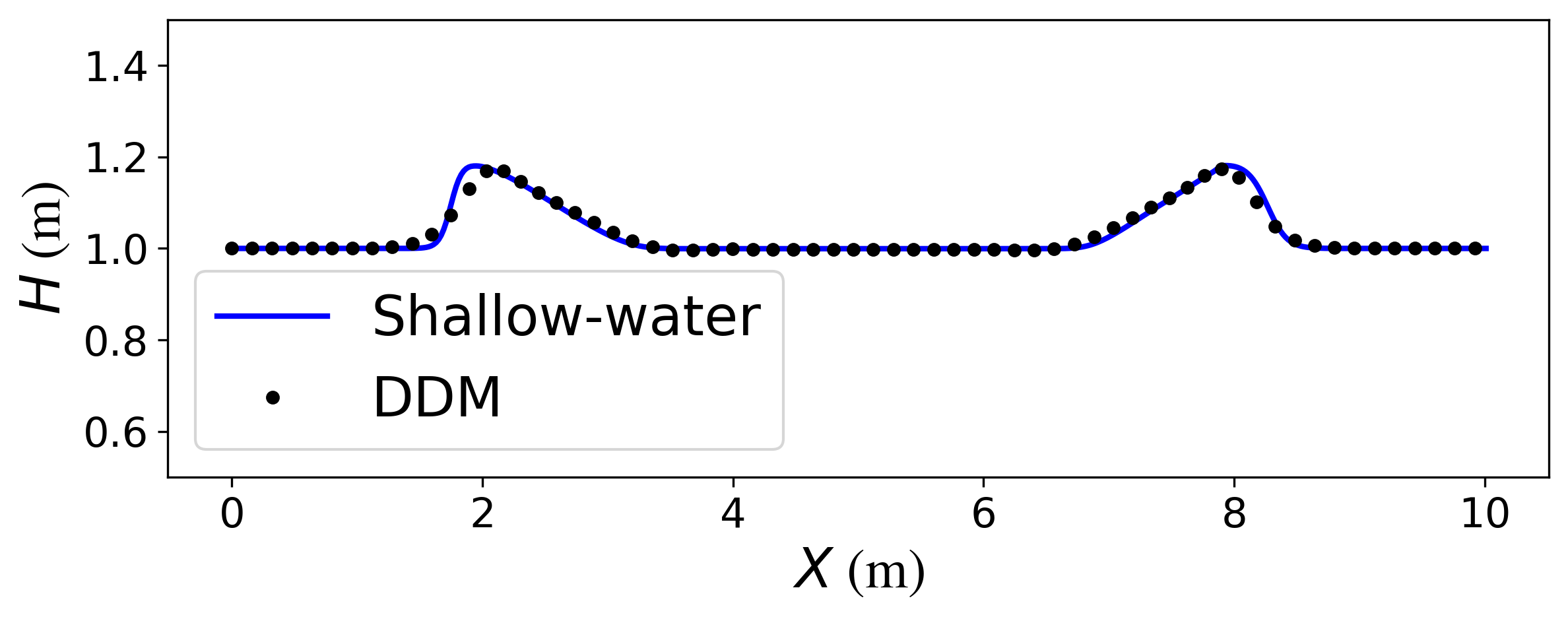}} 
    \subfloat[$t=3.5$s]{\includegraphics[width=2.5in]{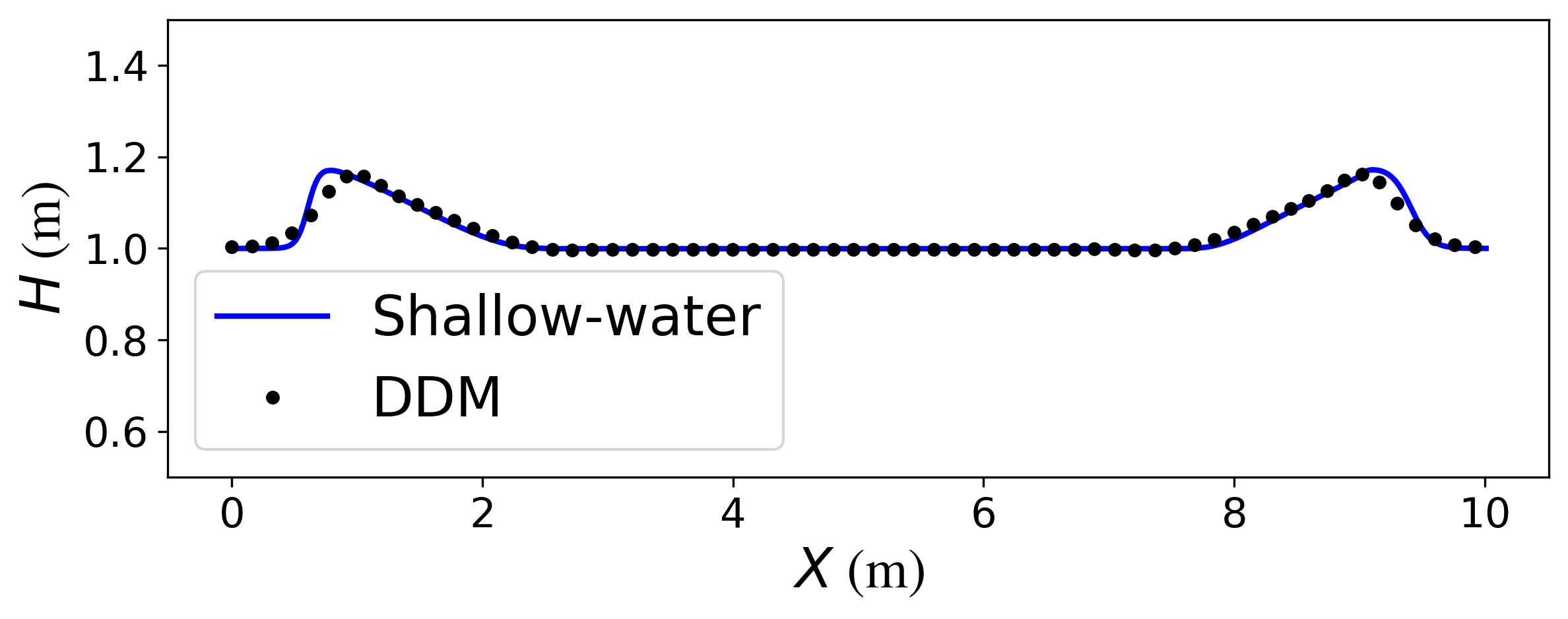}} 
    \caption{Water hump problem: Comparison of the computed height function obtained by the DDM against a finite volume solution to the shallow water equations.}
    \label{fig:water_hump}
\end{figure}

\begin{figure}
    \centering
    \includegraphics[width=0.55\textwidth]{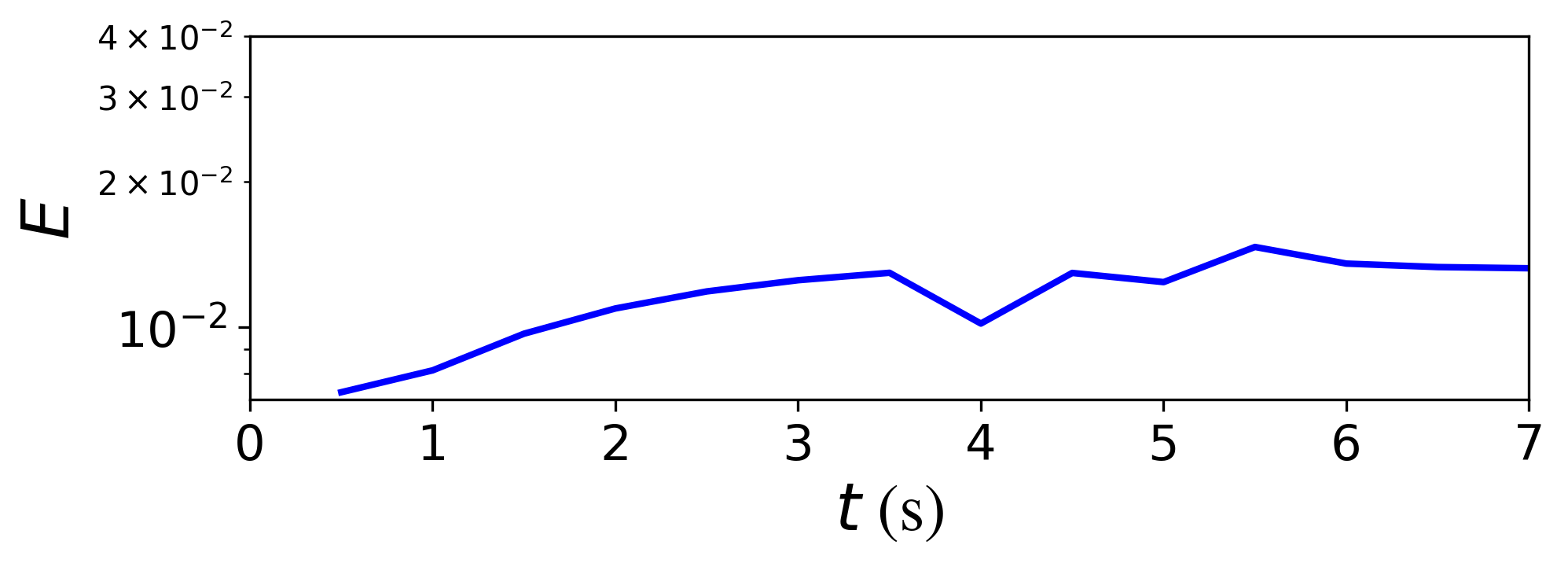}
    \caption{Water hump problem: Evolution of $L_2$ error in height of the DDM solution, considering the height from the FV solution to the shallow water equations as a reference solution.}
    \label{fig:L2_error}
\end{figure}

The discrete droplet solution to the height function is compared with the shallow-water solution in Figure \ref{fig:water_hump} at various time instants of the simulation. The figure shows that both models produce very similar results. 
Figure \ref{fig:L2_error} shows the evolution of the relative $L_2$ error in the DDM solution with time. Here, the FVM solution to the shallow water model is taken as the reference solution. The error at a given time instant is thus measured as 
\begin{equation}
    E = \frac{\vert H_{\text{FVM}}-H_{\text{DDM}}\vert}{\vert H_{\text{FVM}} \rvert}
\end{equation}
Here, $H_{\text{FVM}}$ is the finite-volume method computed height to the shallow water model, and $H_{\text{DDM}}$ is the solution of the proposed discrete droplet method. This relative difference between the solutions of the two models remains bounded under  $1.4 \%$, as the simulation time progresses, as seen in the figure.

\subsection{1-D dam-break problem}
We now consider another 1-D dam-break test case that is a classical problem for benchmarking free-surface and thin-film flows, motivated by \cite{leveque2002finite,saucedo2017new,basic2022lagrangian}. Here, the DDM is tested for an initial condition with a discontinuous height function. The acceleration due to gravity $g$ is taken to be $9.81 \text{m/s}^2$. For the domain $0 \le x \le 1000$, with a zero initial velocity, and initial height function given by
\begin{equation}
    H(x) =
        \begin{cases} 
            10, & \text{for } x<500\\
            1. & \text{for }  x>500
         \end{cases}
\end{equation}

\begin{figure}
    \centering
    \subfloat[Initial height function for the DDM, which is a smoothed approximation of the discontinuous initial condition.]{
    \includegraphics[width=0.45\textwidth]{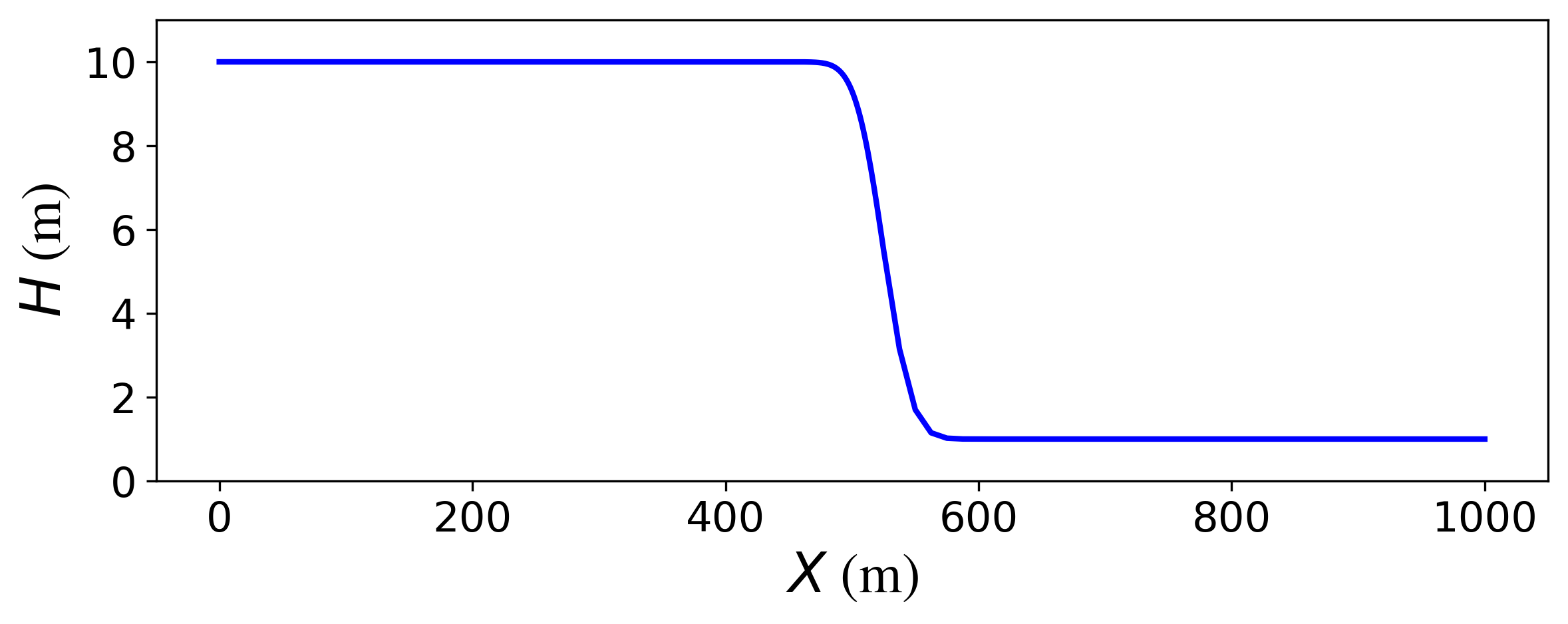}}
    \phantom{Ab}
    \subfloat[Comparison of computed height function against a shallow water solver at time $t=30$s.]{
    \includegraphics[width=0.45\textwidth]{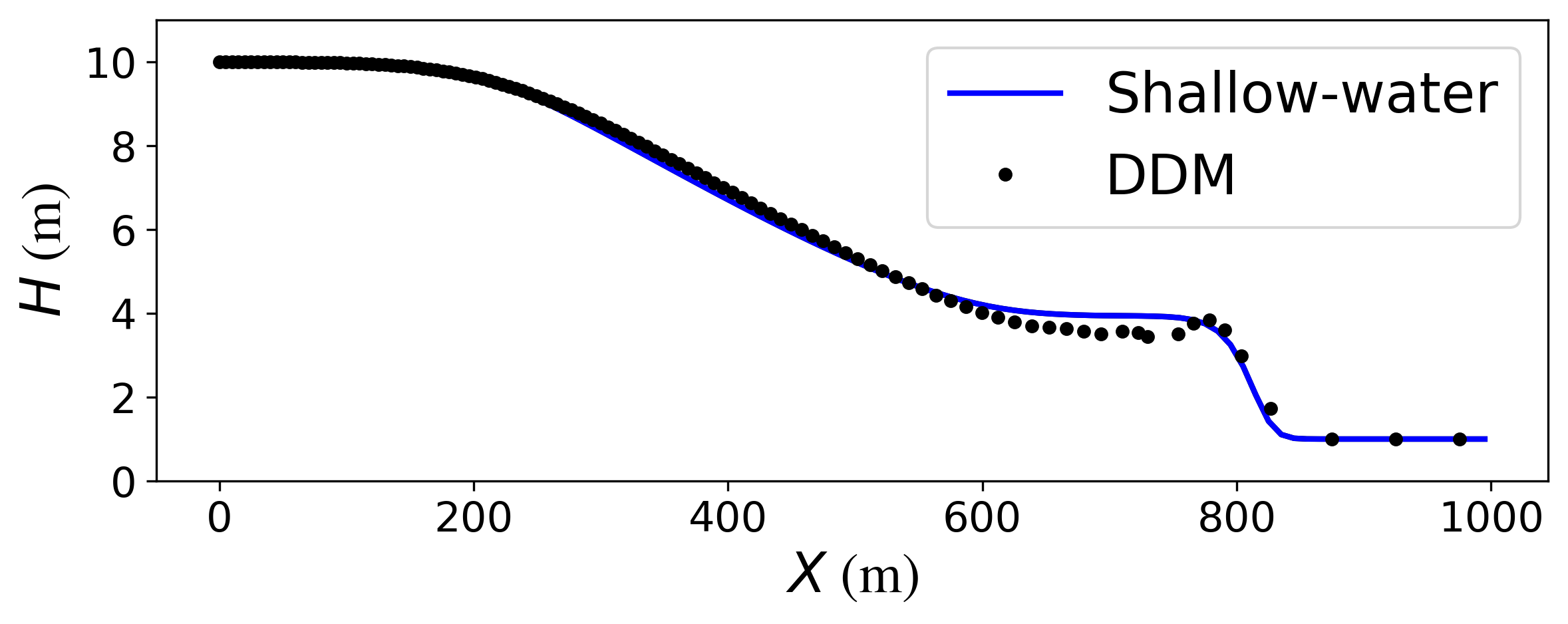}
    }
    \caption{1-D dam-break problem: Initial condition and numerical result.}
    \label{fig:dam}
\end{figure}

The initial droplet distribution for the DDM to match this initial condition is computed using the procedure laid out in Sec \ref{method:initial}. The nature of the inverse problem makes is quite challenging to capture the height discontinuity in the initial condition. Instead, the resultant initial height of the DDM model is a smooth approximation to the discontinuous initial condition, as shown in Figure \ref{fig:dam}(a). 

The solutions of the DDM and the finite-volume shallow-water solver are compared in Figure \ref{fig:dam}(b) at $t=30$s. Both methods produce similar results, however the DDM produces minor oscillations near the location of the discontinuity. The appropriate improvements to the DDM necessary to better capture discontinuous solutions remains an open question beyond the scope of the present work.

\subsection{Sliding thin-film of constant height}

\begin{table}[]
    \centering
    \caption{Sliding thin film of constant height: Parameters of the simulation.}
    \begin{tabular}{cc}
    \hline 
    Parameter     & Value  \\ \hline 
    Viscosity ($\eta$)     & $0.001 \text{ Ns/m}^2$\\
    Density ($\rho$) & $1000 \text{ kg/m}^3$ \\
    Droplet diameter ($d$) & $0.0075 \text{ m}$ \\
    Smoothing length ($h$) & $0.1 \text{ m}$ \\
    Acceleration due to gravity ($g$) & $10 \text{ m/s}^2$ \\ \hline
    \end{tabular}

    \label{tab:c3_parameters}
\end{table}

\begin{figure}
    \centering
    \includegraphics[width=0.75\textwidth]{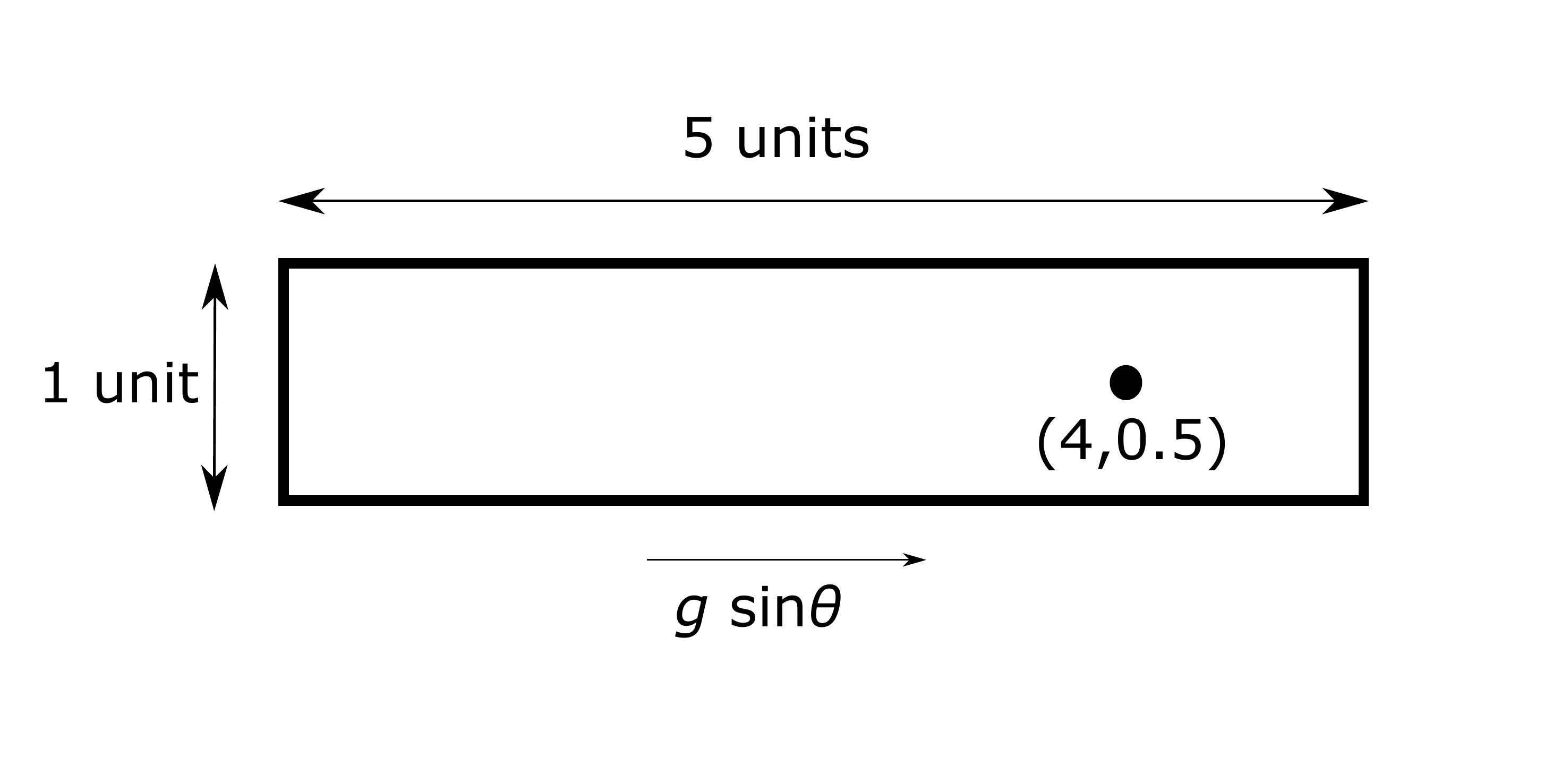}
    \caption{Sliding thin film of constant height: Top-view of the computational domain along which the film slides. The actual plate is at an incline of $\theta = 30^o$. The point where the numerical solution is compared against the analytical one is marked. }
    \label{fig:inc_plane}
\end{figure}

Having compared the solution of the simplified 1-D model on benchmark cases, we now move on to validate the full 2-D DDM model. We consider a test case consisting of a fluid film of constant height sliding over an inclined plane. Thus, the surface considered is 2-D and flat and serves as a good starting point for the validation of the full model. The computational domain, illustrated in Figure \ref{fig:inc_plane}, consists of a plate that is 5 units in length and 1 unit in width, and inclined at an angle of $\theta = 30^o$. The parameters used in this simulation are tabulated in Table \ref{tab:c3_parameters}.

The forces that act on each infinitesimal fluid element are gravitational and the viscous force. To determine an analytical solution, we assume the height of the film to be constant as the film slides down the plane. This represents fluid flowing down an infinite plane where fluid will not collect on one end. Since the height of the film is constant, there are no hydrostatic pressure forces due to the gradient of height. We further note that for an initially stationary fluid layer, the velocity will be non-zero only in the direction of the effective gravity on the inclined plate. Using Eq.~\eqref{eq:2d_momentum}, the analytical solution for this problem that is governed by the following equation.
\begin{equation}
    \frac{d V_{\text{drop}} }{dt} + \frac{\eta_{\text{drop}} V_{\text{drop}}}{\rho_{\text{drop}} H^2} = g \sin \theta 
\end{equation}
where $V_{\text{drop}}$ is the velocity in the direction along the effective gravity (see Figure~\ref{fig:inc_plane}), $g = \| \vec{g} \|$ is the acceleration due to gravity with $g \sin \theta$ the component along the inclined plate,  and $\theta$ is the angle of inclination of the plane.
The analytical solution for the velocity obtained from the method of integrating factors is given by
\begin{equation}
    V_{\text{drop}} = g \sin \theta \frac{\rho_{\text{drop}} H^2}{\eta_{\text{drop}}}\left(1-\exp\left[-\frac{\eta_{\text{drop}} t}{\rho_{\text{drop}} H^2}\right]\right)
\end{equation}

In this test case, a uniformly spaced  droplet distribution is used at the beginning to achieve a constant initial film height $H=2.87e-4$ m. 

\begin{figure}
    \centering
    \includegraphics[width=3in]{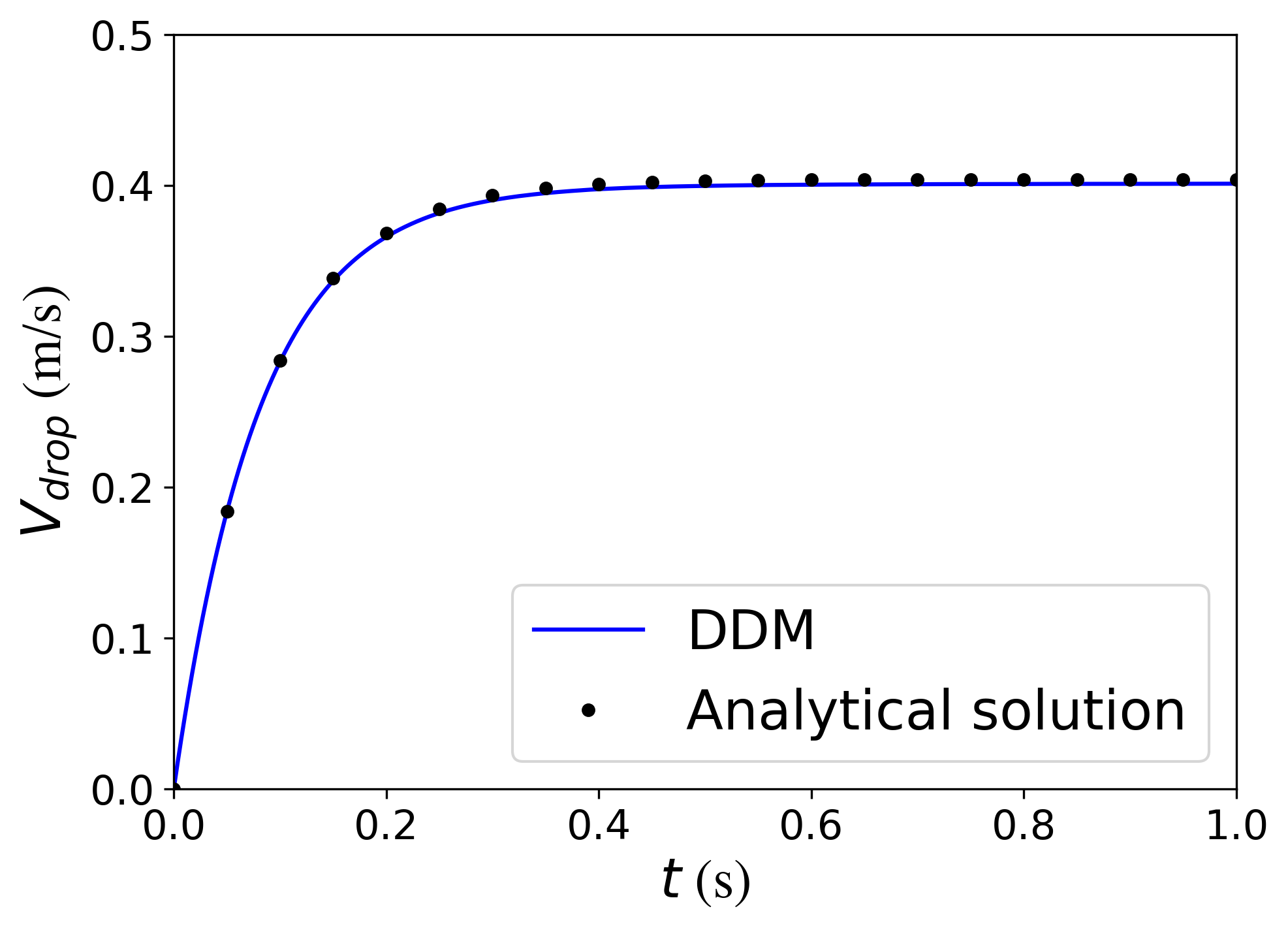}
    \caption{Sliding thin film of constant height: comparison of the numerical velocity at the reference location against the analytical solution.}
    \label{fig:inc_plane_results}
\end{figure}

We compare the numerical velocity to the analytical one at the location $(4,0.5)$ in a plate coordinate frame (see Figure~\ref{fig:inc_plane}). Since a droplet may not be present at that exact location, the numerical velocity is interpolated at that location from all droplets in it's neighbourhood. Figure \ref{fig:inc_plane_results} compares the time evolution of the velocity obtained from the DDM against the analytical solution. The figure shows that the numerical result closely matches the analytical one. 

\subsection{Thin-film in a hemispherical bowl}

\begin{table}[]
    \centering
    \caption{Thin film in a bowl: Parameters of the simulation}

    \begin{tabular}{cc}
    \hline 
    Parameter     & Value  \\ \hline 
    Viscosity ($\eta$)     & $0.01$ Ns/m$^2$\\
    Density ($\rho$) & $1000$ kg/m$^3$ \\
    Droplet diameter ($d$) & $0.001$ m \\
    Smoothing length ($h$) & $0.008$ m \\
    Acceleration due to gravity ($g$) & $10$ m/s$^2$ \\ \hline
    \end{tabular}
    \label{tab:c4_parameters}
\end{table}

\begin{figure}
    \centering
    \includegraphics[width=0.5\textwidth]{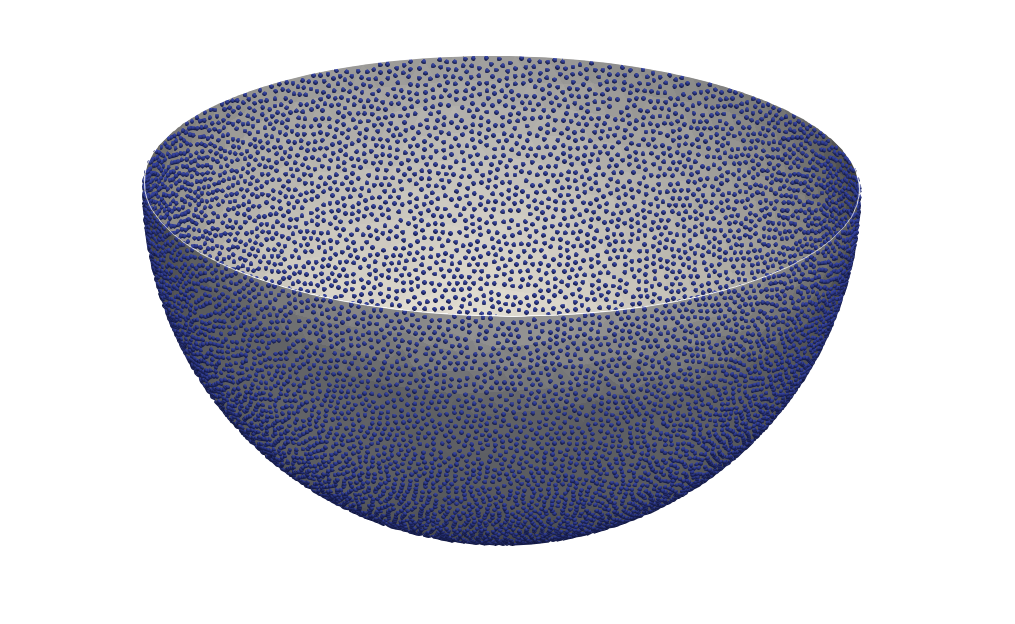}
    \caption{Thin film in a bowl: Initial setup. The hemispherical bowl is initialized with a thin film using scattered droplets.}
    \label{fig:geo_bowl}
\end{figure}

\begin{figure}
    \centering
    \subfloat[$t=0$s]{\includegraphics[width=0.5\textwidth,trim={5cm 0 5cm 0}]{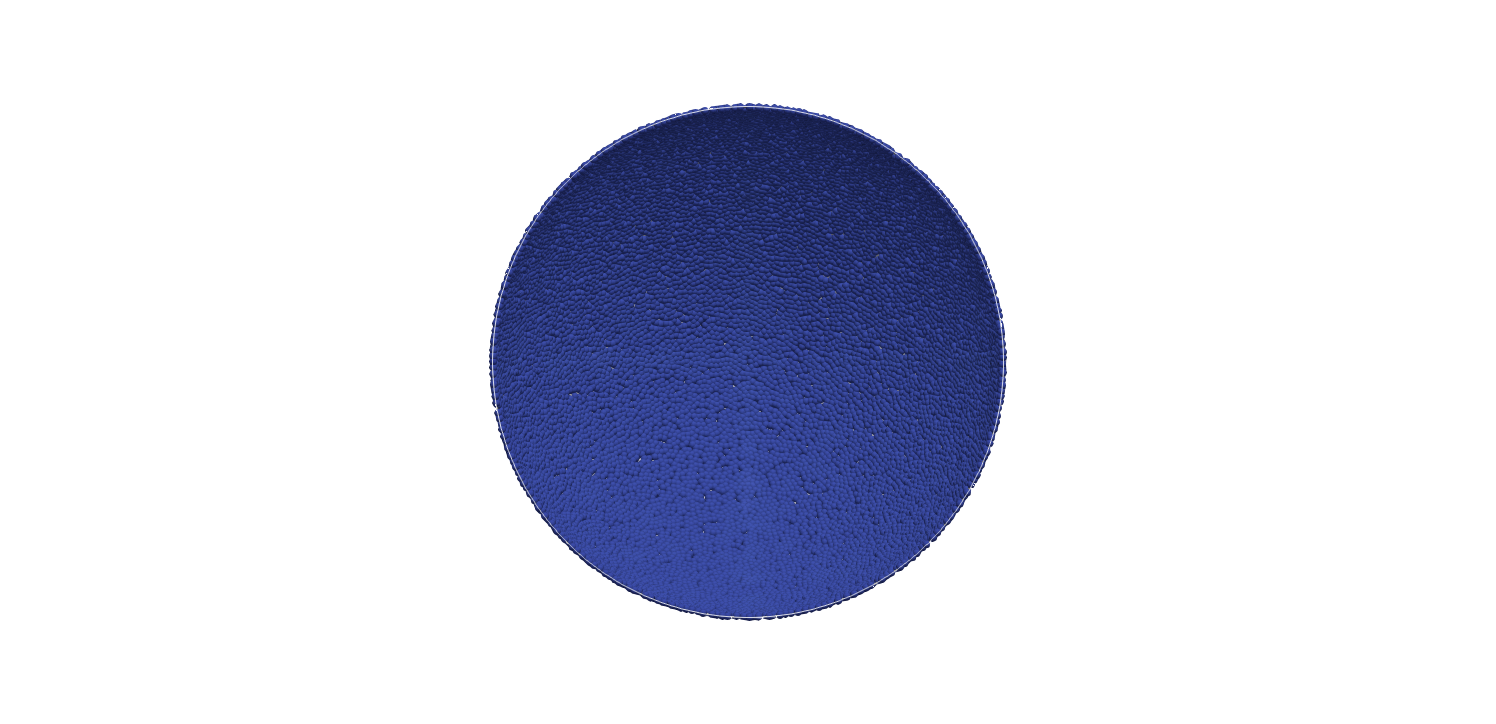}}
    \subfloat[$t=0.5$s]{\includegraphics[width=0.5\textwidth,trim={5cm 0 5cm 0}]{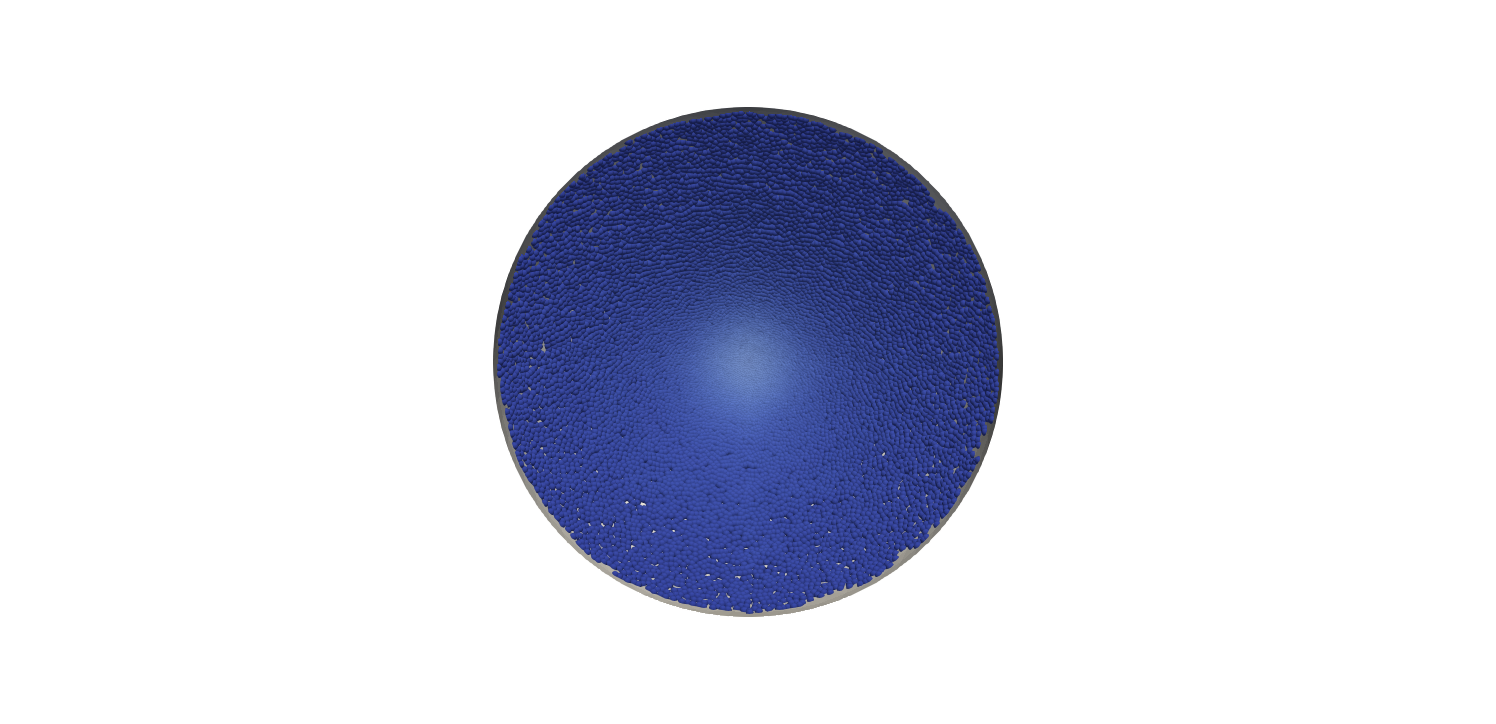}}\\
    \subfloat[$t=1.0$s]{\includegraphics[width=0.5\textwidth,trim={5cm 0 5cm 0}]{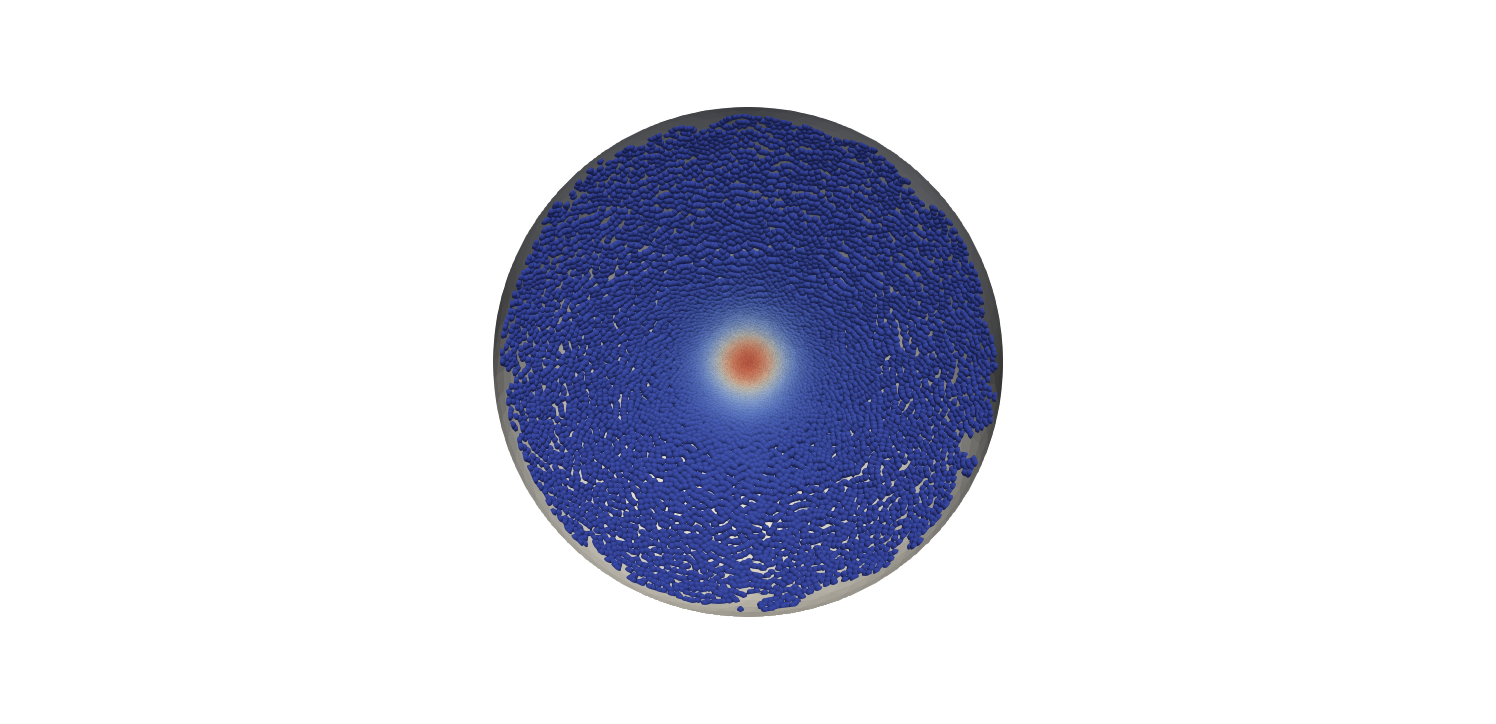}}
    \subfloat[$t=1.5$s]{\includegraphics[width=0.5\textwidth,trim={5cm 0 5cm 0}]{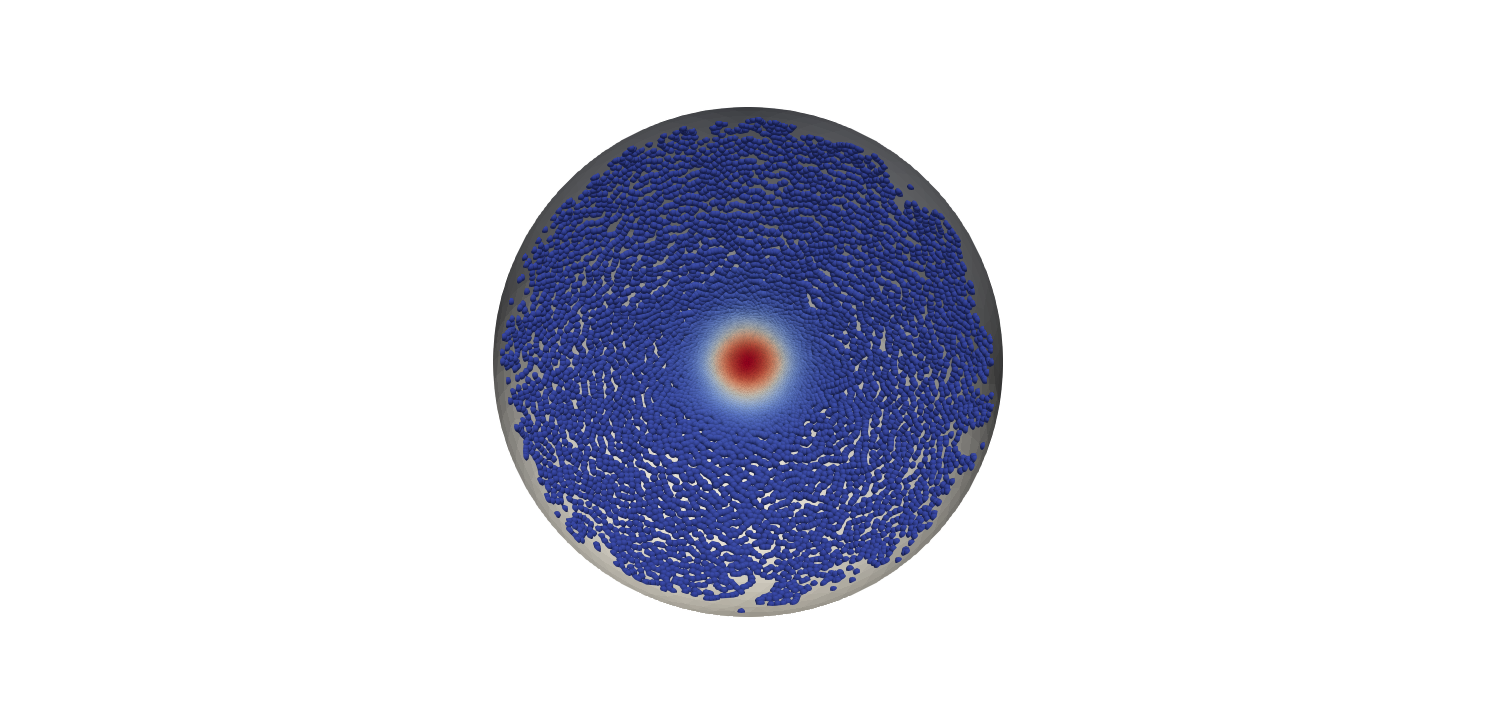}}\\
    \caption{Thin film in a bowl (top view): Evolution of the film as it collapses due to gravity. The colour represents the film height $H$. See Figure~\ref{fig:geo_bowl} for a side profile of the domain. } 
    \label{fig:bowl_frames}
\end{figure}

In this test case, a hemispherical bowl, as shown in Figure \ref{fig:geo_bowl}, is considered. Building on the previous case, this case involves a curved 2-D surface with a thin film. The parameters used in this simulation are tabulated in Table \ref{tab:c4_parameters}. The bowl is initialized with a thin film of constant height at rest, that collapses into the center of the bowl under the influence of gravity. Unlike the previous test case, both viscous and pressure gradient forces play a role in the evolution of the film. As it collapses, the height of the film at the bottom of the bowl increases, as seen in Figure \ref{fig:bowl_frames}. This figure shows the top view of the bowl and the coloured contours of the computed height function at different instances of time. 
%

In each of the earlier test cases, we considered a fixed droplet diameter. We now examine how the choice of the droplet diameter affects the solution. A finer droplet diameter can be considered as a finer resolution of the DDM. Thus, the numerical results should converge with reducing diameter, which is the consideration of the present test case. 

\begin{figure}
    \centering
    \includegraphics[width=0.6\textwidth]{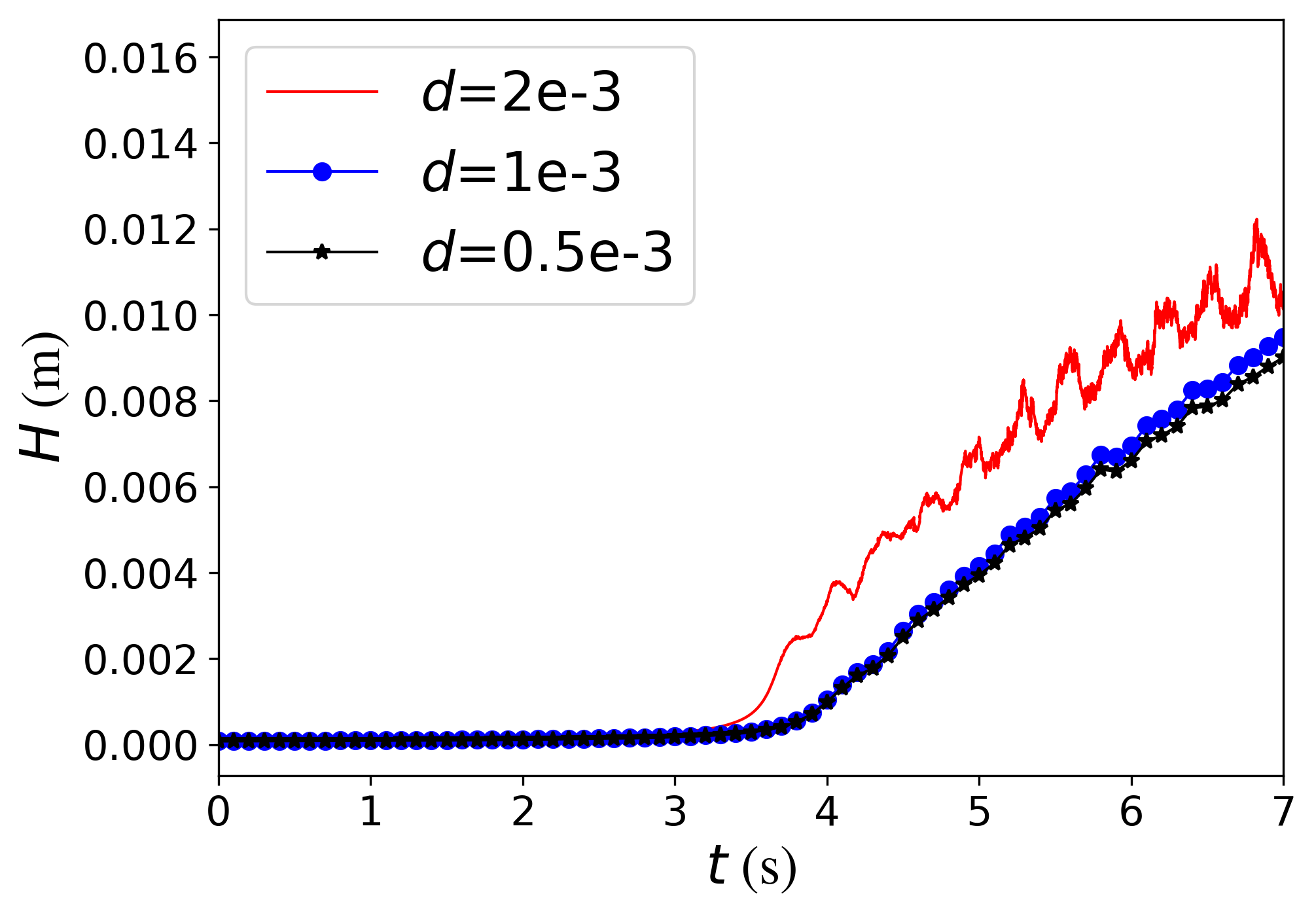}
    \caption{Thin film in a bowl: Convergence of height function at the bottom of the bowl with varying droplet diameters.}
    \label{fig:height_bowl}
\end{figure}

Three simulations are considered that make use of different diameter of droplets. The droplets are uniformly spaced in each case, such that the initial height of the film is approximately the same. To achieve the same initial film height, lesser droplets of the larger diameter are needed, and hence, they are spaced further apart than the smaller diameter cases. 
The test cases considered are tabulated in Table \ref{tab:dia_cases}. 
The total volume of the fluid in the thin film varies by up to $7 \%$ from that for the largest diameter. Thus, the initial condition in each of the three test cases is only approximately the same, and not exactly so. 
\begin{table}[]
    \centering
    \caption{Thin film in a bowl:  Cases considered with different droplet diameters. $N$ is the total number of the droplets in the domain. }
    \begin{tabular}{ccc}
        \hline
        \hline
         Simulation & Diameter, $d$ (m) & $V=\frac{N \pi d^3}{6}$ (m$^3$)  \\ \hline
         1 & $2e-3$ & $5.6e-6$ \\
         2 & $1e-3$ & $5.3e-6$ \\
         3 & $0.5e-3$ & $6e-6$ \\
         \hline
    \end{tabular}
    \label{tab:dia_cases}
\end{table}
Figure \ref{fig:height_bowl} shows a plot that compares the height evolution at the bottom of the hemispherical bowl for each of the three simulations. It is seen that the height closely follows the same trend in the three simulations and evolves more smoothly in the case of smaller diameter of droplets (Simulations $2$ and $3$). Additionally, it is noted that reducing the droplet size does not lead to identical converged solutions since the droplets present on the bowl at $t=0$ only approximately represent the same film height. The initial differences in the film height manifest in the height evolution during the course of the simulation.

\subsection{Asymmetric Dam-break problem}

\begin{table}[]
    \centering
    \caption{Asymmetric Dam-break: Parameters of the simulation}
    \begin{tabular}{cc}
    \hline 
    Parameter     & Value  \\ \hline 
    Viscosity ($\eta$)     & 0.001 Ns/m$^2$\\
    Density ($\rho$) & 1000 kg/m$^3$ \\
    Droplet diameter ($d$) & 2.0 m \\
    Smoothing length ($h$) & 2.2 m \\
    Acceleration due to gravity ($g$) & 10 m/s$^2$ \\ \hline
    \end{tabular}
    \label{tab:c5_parameters}
\end{table}

\begin{figure}
    \centering
    \includegraphics[width=0.4\textwidth]{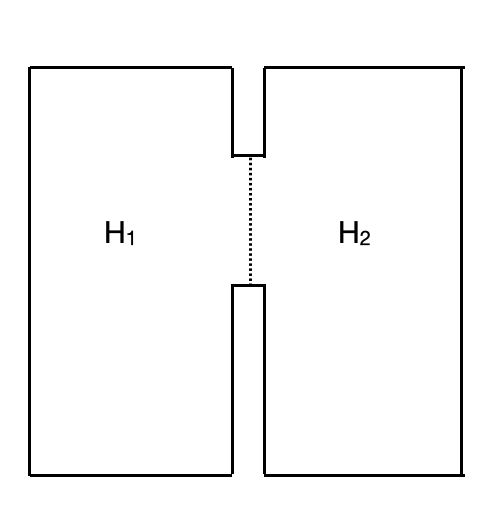}
    \caption{Asymmetric Dam-break: An illustration of the computational domain. $H_1$ and $H_2$ are the initial heights of the fluid films in each sub-domain, with $H_1>H_2$. Fluid thus flows from the left sub-domain to the right one.}
    \label{fig:adam_illus}
\end{figure}

This test case is adapted from \cite{mingham1998high} and is presented to illustrate the performance of the full model in the presence of a discontinuity. It involves the flow of fluid from one side of a dam of height $H_1=10$m, to the other of height $H_2=1$m. Figure \ref{fig:adam_illus} shows an illustration of the domain. Table \ref{tab:c5_parameters} presents the parameters used in this test case. The initial discontinuity in the height function drives the flow from left to right.  

\begin{figure}
    \centering
    \subfloat[3-D solver, $t=3s$]{\includegraphics[width=0.49\textwidth]{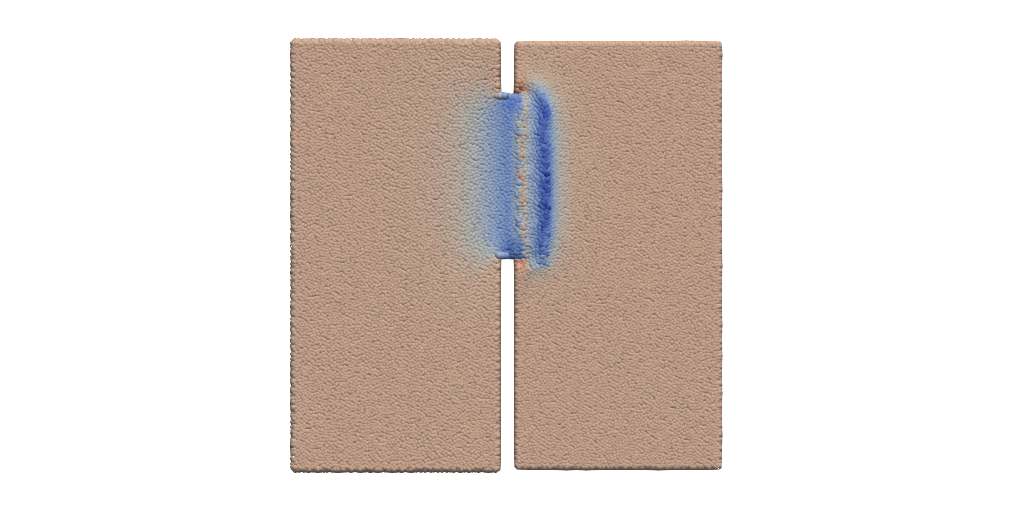}}
    \subfloat[3-D solver, $t=5s$]{\includegraphics[width=0.49\textwidth]{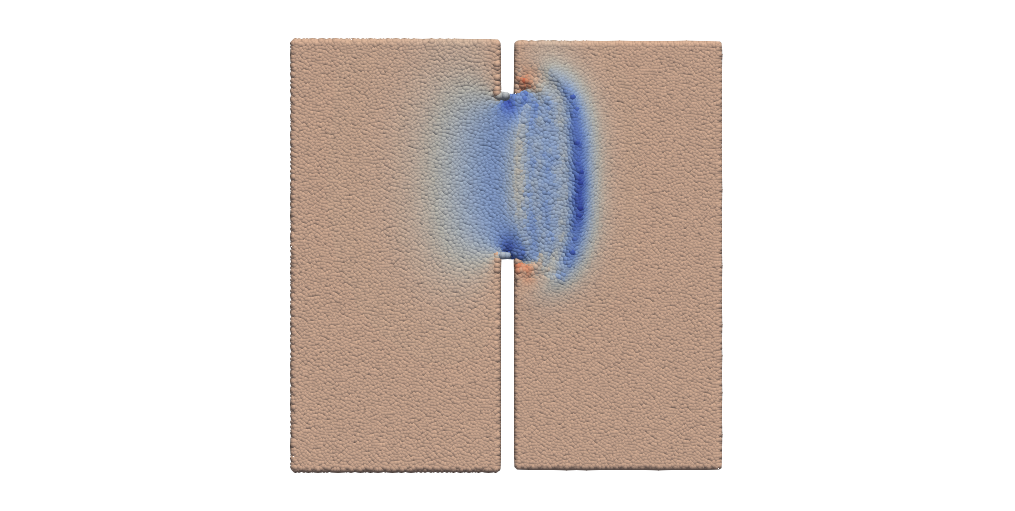}}\\
    \subfloat[DDM, $t=3s$]{\includegraphics[width=0.49\textwidth]{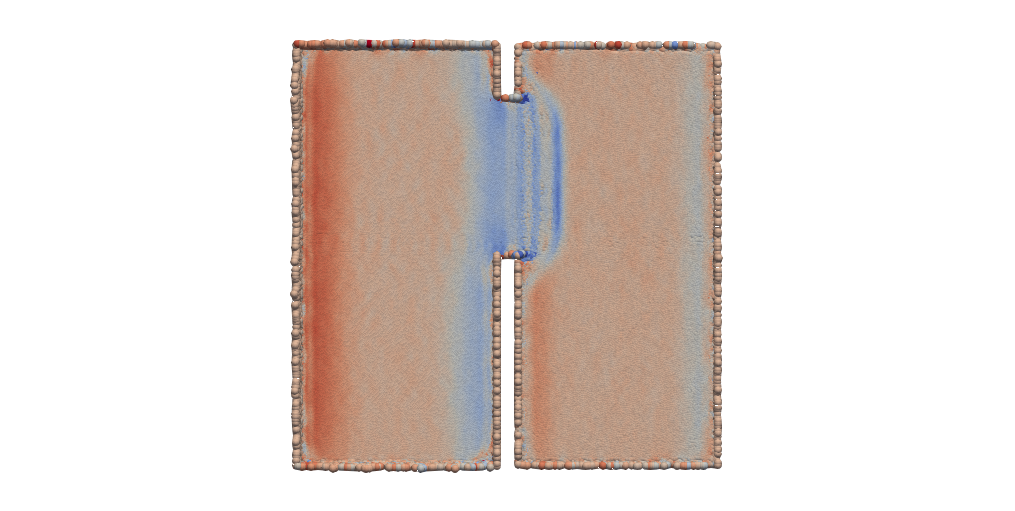}}
    \subfloat[DDM, $t=5s$]{\includegraphics[width=0.49\textwidth]{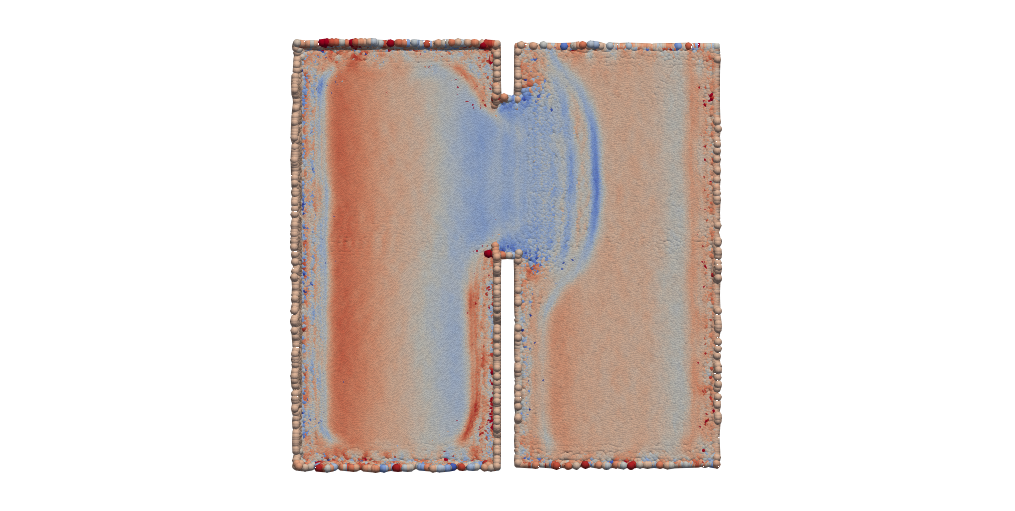}}
    \caption{Asymmetric Dam-break: Comparison of DDM and 3-D Navier--Stokes solver. The colour represents the velocity.}
    \label{fig:asymmetric_dam}
\end{figure}

\begin{figure}
    \centering
    \includegraphics[width=0.5\textwidth]{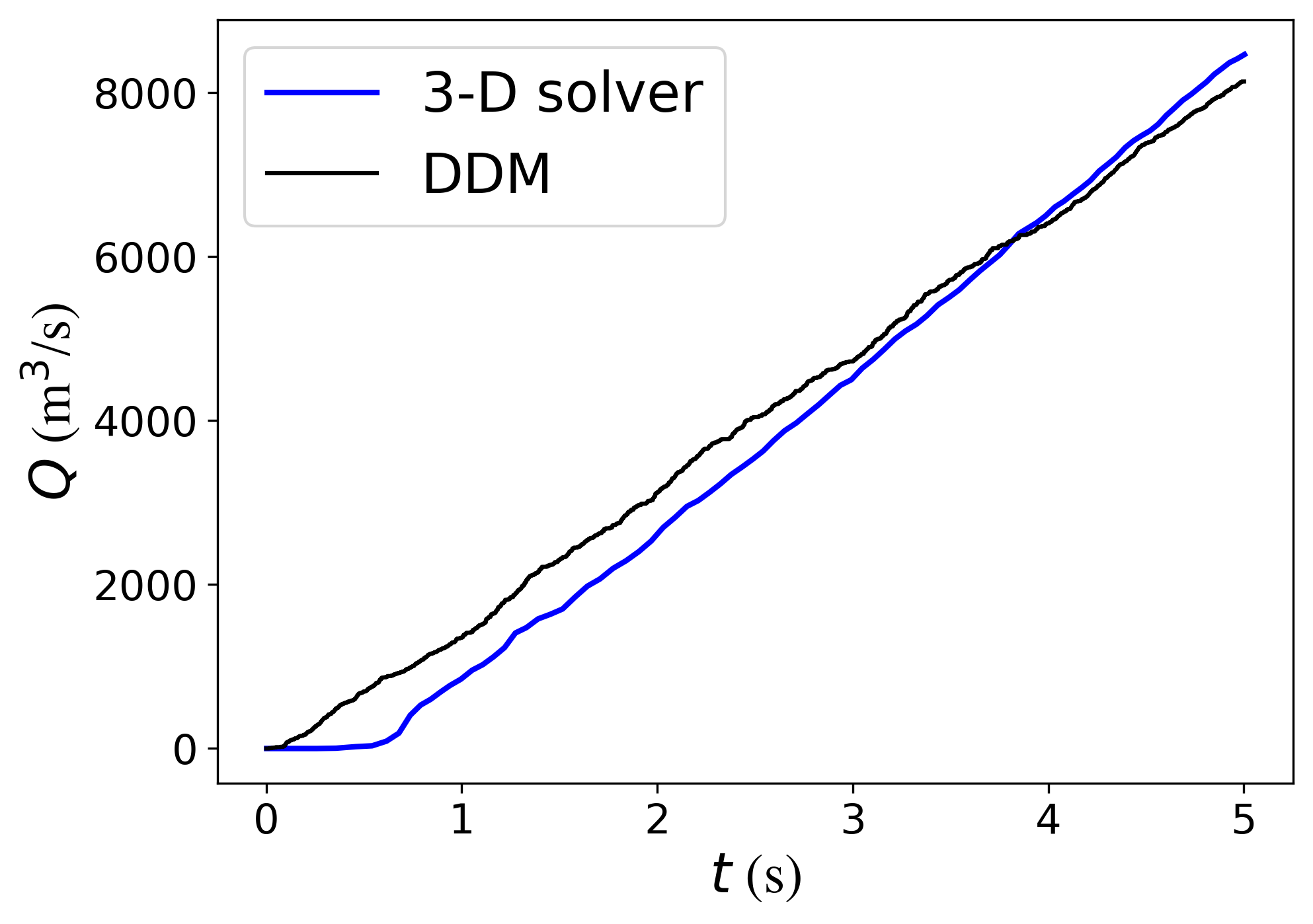}
    \caption{Asymmetric Dam-break: Comparison of cumulative volume fluxes of DDM and 3-D Navier-Stokes solver.}
    \label{fig:adam_mass}
\end{figure}

The test cases are simulated using the DDM as well as a 3-D Navier--Stokes solver. Figure \ref{fig:asymmetric_dam} shows the solution (x component of the velocity) contours at two different instants of time, as solved by the two models. We observe that the wave fronts carrying the liquid from the left sub-domain to the right one match well for the two models. However, there are some differences noted in the two solutions. Firstly, the wave front looks sharper in the case of the 3-D solver. A likely reason is that the 3-D solver captures the initial discontinuity perfectly, while the droplet-solver only approximately captures the jump using a small region of high gradient of height. Secondly, the droplet-solver is seen to have noisier results near the boundaries. This is attributed to the fact that the height function is not constant near the boundaries of the droplet-solver. The non-zero height gradients, therefore, result in the motion of the droplets near the boundaries, which is not observed in the 3-D solver. Figure \ref{fig:adam_mass} reveals a close comparison of the cumulative volume flux of fluid flowing from the left to the right sub-domain for the two models.   


\subsection{Flow over the spoiler of a car}

\begin{table}[]
    \centering
    \caption{Flow over a spoiler: Parameters of the simulation}
    \begin{tabular}{cc}
    \hline 
    Parameter     & Value  \\ \hline 
    Viscosity ($\eta$)     & 0.001 Ns/m$^2$\\
    Density ($\rho$) & 1000 kg/m$^3$ \\
    Droplet diameter ($d$) &  0.002 m \\
    Smoothing length ($h$) & 0.004 m \\
    Acceleration due to gravity ($g$) & 10 m/s$^2$ \\ \hline
    \end{tabular}
    \label{tab:c6_parameters}
\end{table}

We now consider a much more complex application. This test case illustrates a target application of the DDM, consisting of a fluid film forming over a curved surfaces, with no initial film present. The parameters of this test case are tabulated in Table \ref{tab:c6_parameters}. We consider the simulation of rainwater falling on the spoiler of a car, and forming a thin film over it. We simulate the formation and evolution of the film, followed by water leaving the spoiler at the edges. The spoiler geometry used in this simulation is an actual in-production spoiler used in the automotive industry. 

\begin{figure}
    \centering
    \includegraphics[width=0.6\textwidth]{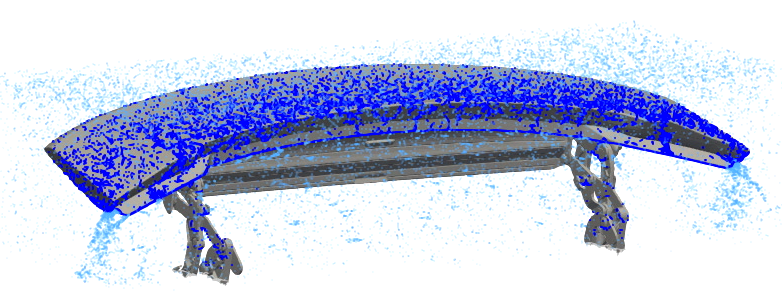}\\
    \includegraphics[width=0.6\textwidth]{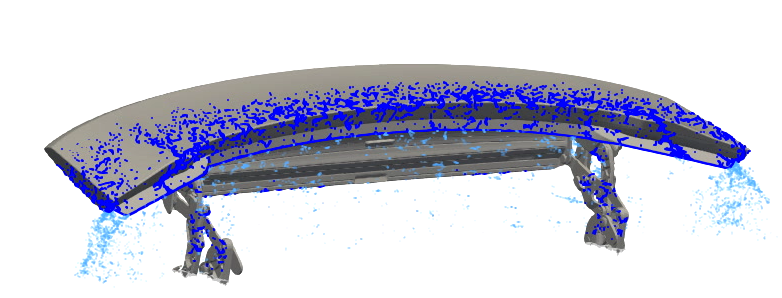}\\
    \includegraphics[width=0.6\textwidth]{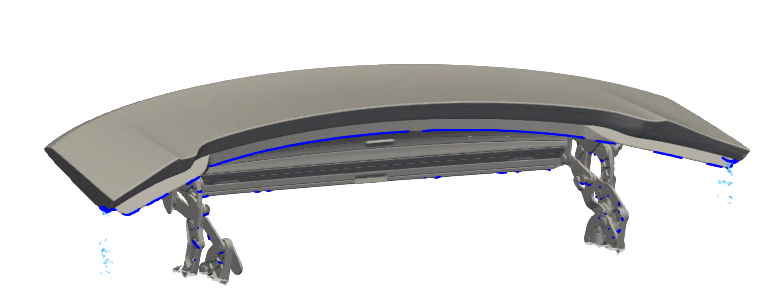}
    \caption{Flow over the spoiler of a car using the discrete droplet method. (Top) Rain falling on the spoiler, and forming a thin film over it, (middle, bottom) water collecting in some regions as the rain stops. }
    \label{fig:spoiler}
\end{figure}

The simulation starts with no fluid present in the domain, neither in free-flight, nor as a thin film. At $t=0$s, droplets in free-flight enter the domain from an inflow plane over the spoiler. As seen in Figure  \ref{fig:spoiler},  as droplets hit the spoiler, they form a film over it. For all droplets hitting the spoiler, the thin-film DDM is used. At $t=5$s, the source of inflow of water is switched off, and all the remaining droplets flow over the edges of the spoiler. A key motivation behind these simulations is to study the locations where fluid collects over the spoiler, and the disruption of water at the edges. 

An equivalent simulation is also performed using the 3-D Navier--Stokes solver of the software MESHFREE, with the same fluid inflow. The mass flux of fluid over the edges of the spoiler is compared from the two models. For this comparison, the mass flux is calculated by computing the amount of fluid passing through a fictitious plane just below the spoiler. This serves not just as an additional validation test case for the thin-film DDM, but also serves to validate the modelling of fluid leaving the thin film, as described in Section~\ref{sec:Leave}.

\begin{figure}
    \centering
    \includegraphics[width=0.5\textwidth]{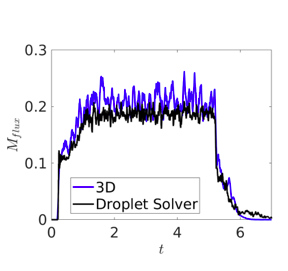}
    \caption{Flow over a spoiler: Comparison of mass flux of water flowing over the edges: DDM  (droplet-solver) vs full a Navier--Stokes solver (3D). }
    \label{fig:massflux}
\end{figure}

A comparison of the mass flux over time is shown in Figure \ref{fig:massflux}. The figure shows that the mass flux computed with both models are very similar, with lesser fluctuations in time in the DDM solver. An important point to note here is that to simulate this with a 3-D resolved method, a very fine resolution was needed to accurately capture the thin layers of fluid. As a result, the thin-film solver was significantly faster than the 3-D one. The 3-D Navier--Stokes solver took 12.5 hours to complete, while running in parallel on $28$ cores. On the other hand, the DDM simulation took only 40 minutes in comparison, while running in parallel on just $10$ cores. This illustrates the computational advantage of using the DDM over the 3-D Navier--Stokes model in capturing such film behaviour. 

\section{Conclusion} \label{sec:conclusions}
A novel \emph{discrete droplet method} is proposed to model the behaviour of thin fluid films over curved surfaces, by tracking moving fluid droplets and their aggregation. The full model that describes thin-film behaviour on surfaces in $\mathbb{R}^3$ is presented, along with a 1-D simplification of the model for validation. Instead of solving an additional PDE for the height function, the proposed model makes use of relative droplet positions and their diameters to reconstruct the height function and its derivatives. The height and its derivatives are used in the computation of the velocities of the droplets, using which, the droplets are moved in a Lagrangian fashion to obtain their locations for the next time step. The proposed model is thoroughly verified and validated by comparing it with other thin film flow solvers, an analytical solution, and with the 3-D Navier--Stokes equations. 

One of the prominent advantages of this method is that since the thin film is modeled as droplets, the method can easily be used for modelling partially wetted surfaces, while restricting the discretization only to the region of presence of the film. Another important advantage is that it can be used to model the capture the formation of a fluid film, when no thin film is present at the beginning of the simulation. The model can also be very easily coupled with free-flight droplet models. 

Several open questions need to be addressed, which form the direction of our future work. The proposed DDM, like other existing thin-film models, has to obey the thin-film assumption that the film-depth is much smaller than the lateral dimensions of the film. Therefore, in cases where a film is just forming, there may be situations where the number of droplets would be too few to obey the thin-film assumption. One of the future directions of research could be to establish a criterion based on the thin-film assumption, to ascertain when the droplet aggregation on a surface is sufficient to form a thin-film, or whether a simplified model should be used. Another open question is to develop better approaches to solve the inverse problem to match initial conditions for the height function, which is especially problematic near boundaries and discontinuities. Extensions of the present work include the use of higher order SPH derivative computation for the height function in order to achieve better accuracy, the coupling of the proposed thin film solver with 3-D Navier--Stokes models, and the incorporation of other physical phenomena relating to thin-films such as evaporation and freezing.

\section*{Acknowledgments}
Anand S Bharadwaj would like to thank the Fraunhofer-Gesellschaft for the financial support provided. Pratik Suchde would like to acknowledge support from the European Union's Horizon 2020 research and innovation program under the Marie Sklodowska-Curie Actions Grant Agreement No. 892761 “SURFING”. Stéphane P.A. Bordas received funding from the European Union’s Horizon 2020 research and innovation programme under grant agreement No 811099 TWINNING Project DRIVEN for the University of Luxembourg. The authors would like to thank Porsche AG for providing the spoiler geometry. 

\appendix
\section{Derivation of Gaussian kernel function for SPH formulation} \label{sec:app1}
The kernel function appearing in the 2-D SPH formulation for the height Eq.~\eqref{eq:height_2dd} , needs to satisfy the consistency condition 
\begin{equation}
\int \int_{\mathbb{R}^2}  W(\vec{x}-\vec{x}_0) dA = 1
\end{equation}

We note here that a locally 2-D neighbourhood is assumed. Without the loss of generality, assume $\vec{x}_0=\vec{0}$ for notational convenience. We assume the form of the kernel function as
\begin{equation}
 W(\vec{x}) = K \exp\left[ -\alpha \frac{\|\vec{x}\|^2}{h^2}\right]
\end{equation}
where $K$ is the coefficient to be determined, $\alpha$ is a scalar constant and $h$ is the smoothing length or interaction radius.
Switching to polar coordinates, we get
\begin{equation}
 W(\vec{r}) = K \exp \left[ -\alpha \frac{r^2}{h^2}\right]
\end{equation}

Substituting in the integral over $\mathbb{R}^2$, 
\begin{equation}
\int_0^{2\pi} \int_0^{\infty}  K \exp\left[ -\alpha \frac{r^2}{h^2}\right] r dr d\theta = 1
\end{equation}

\begin{equation}
\frac{Kh^2}{2\alpha} \int_0^{2\pi} d\theta  \int_0^{\infty}   \exp\left[ -\alpha \frac{r^2}{h^2}\right] d(\alpha r^2/h^2) = 1
\end{equation}

\begin{equation}
\frac{Kh^2}{2\alpha} 2 \pi \left(- \exp\left[ -\alpha \frac{r^2}{h^2}\right] \right) \Bigg \vert _0 ^\infty = 1
\end{equation}

\begin{equation}
\frac{Kh^2}{2\alpha} 2 \pi (1) = 1
\end{equation}

\begin{equation}
K = \frac{\alpha}{\pi h^2}
\end{equation}

Thus, the kernel function is 
\begin{equation}
 W(\vec{x}) = \frac{\alpha}{\pi h^2} \exp\left[ -\alpha \frac{\|\vec{x}\|^2}{h^2}\right]
\end{equation}

\section{Alternate derivation for height function of the discrete droplet solver} \label{sec:app2}

\begin{figure}
    \centering
    \includegraphics[width=0.8\textwidth]{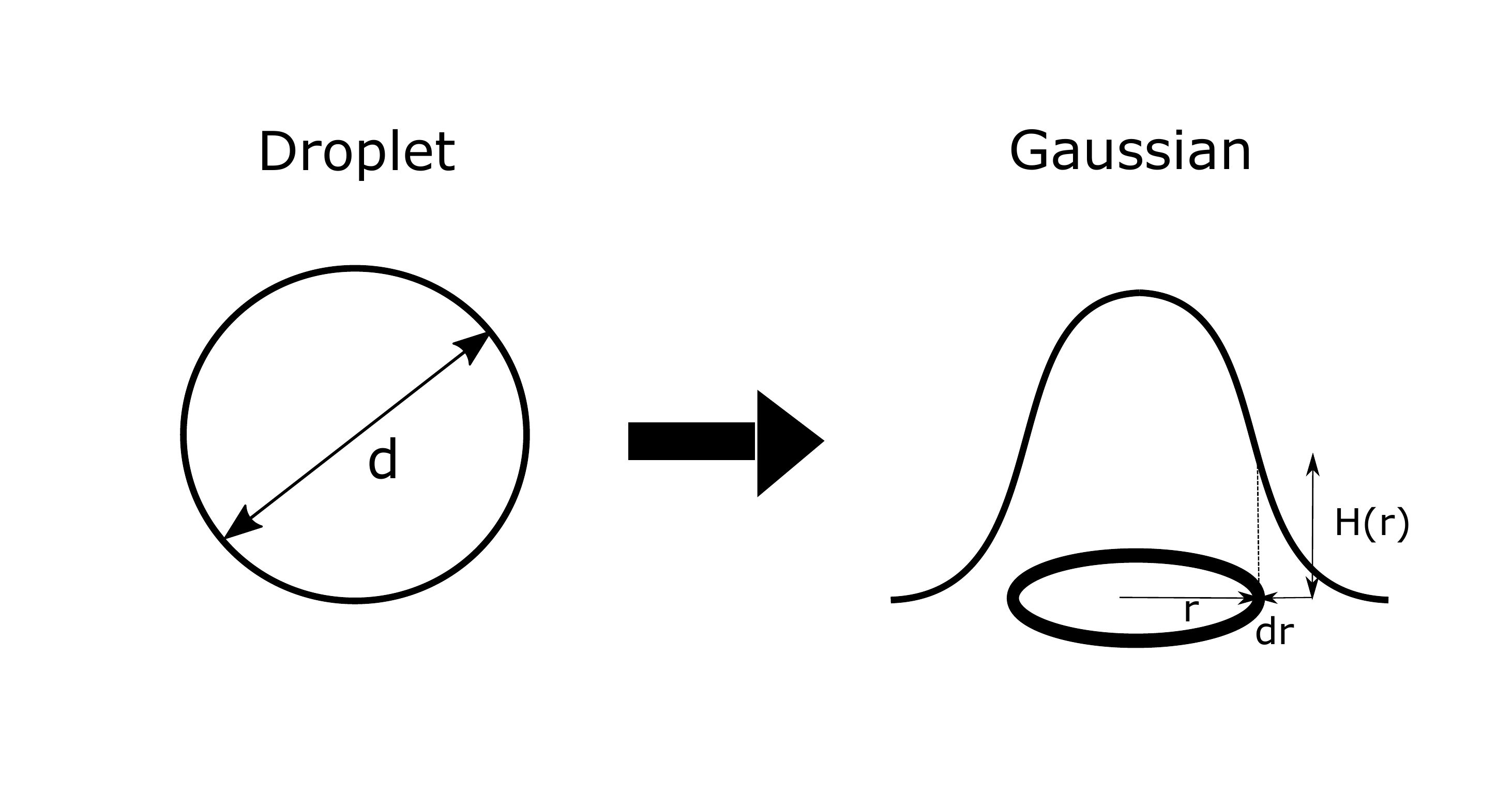}
    \caption{Droplet volume smeared out over a Gaussian shape.}
    \label{fig:droplet_gaussian}
\end{figure}
Here, an alternate approach to arrive at the height function as shown in Eq.~\eqref{eq:height_2dd}, is discussed. Each spherical droplet is associated with a Gaussian kernel function, as illustrated in Figure \ref{fig:droplet_gaussian}. Assume that this function centred at the origin, is of the form
\begin{equation}
    H(r) = K \exp\left(-\alpha\frac{r^2}{h^2}\right) 
\end{equation}
where $r$ is the distance from origin, $K$ is to be determined, $\alpha$ is a scalar constant and $h$ is the smoothing length. 

As per the conservation of mass (or volume, since density is taken constant), the volume of the droplet should be equal to the volume under the Gaussian function.  Therefore, 
\begin{equation}
    \frac{\pi d^3}{6} =  \int \int_A H(r)  dA = \int_{0}^{\infty}  K \exp\left(-\alpha\frac{r^2}{h^2}\right)  2 \pi r dr 
\end{equation}

This may be rewritten as
\begin{equation}
\frac{\pi d^3}{6} = Kh^2 \pi \int_{0}^{\infty}   \exp\left(-\alpha\frac{r^2}{h^2}\right)  d(r^2/h^2)
\end{equation}
\begin{equation}
\frac{\pi d^3}{6} = -\frac{Kh^2}{\alpha} \pi  \exp\left(-\alpha\frac{r^2}{h^2}\right) \Bigg \vert_0^\infty
\end{equation}

This gives, 
\begin{equation}
   K = \frac{\alpha d^3}{6h^2}
\end{equation}

Using the above result, the Gaussian at a given droplet location is
\begin{equation}
    H(r) = \frac{\alpha d^3}{6h^2} \exp\left(-\alpha\frac{r^2}{h^2}\right) = \frac{C^*}{h^2} \exp\left(-\alpha\frac{r^2}{h^2}\right) V
\end{equation}
where $C^* = \alpha/\pi$ and $V = \pi d^3/6$ is the volume of the droplet.

Now, expressing the height function at a droplet $i$ as a summation of the Gaussians of the neighbouring droplets, we get
\begin{equation}
    H_i = \sum_{j \in S_i}  \frac{C^*}{h_j^2} \exp\left(-\alpha\frac{\|\vec{x_j}-\vec{x_i}\|^2}{h_j^2}\right) V_j
\end{equation}
which is the same as the expession in Eq.~\eqref{eq:height_2dd}. 
Here, \\
$H_i$ is the height of the film at a droplet $i$, \\
$j$ is the summation index for the droplets in the neighbourhood of droplet $i$, \\
$\|\vec{x_i}-\vec{x_j}\|$ is the distance between droplets $i$ and $j$.

\bibliographystyle{unsrt}

\bibliography{main.bib}

\end{document}